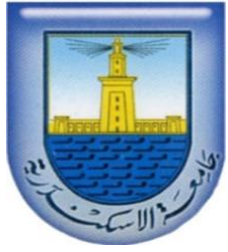

ALEXANDRIA UNIVERSITY

FACULTY OF ENGINEERING

ELECTRICAL ENGINEERING DEPARTMENT

# TECHNIQUES FOR COOPERATIVE COGNITIVE RADIO NETWORKS

A thesis submitted in partial fulfillment for the degree of

## Master of Science

By

**Ramy Mohamed Amer Ghanem**
B.Sc., Faculty of Engineering, Alexandria University, 2010

Supervised by

**Prof. Dr. Mohamed Amr Mokhtar**
Department of Communications and Electronics
Faculty of Engineering, Alexandria University

**Prof. Dr. Amr Ahmed El-Sherif**
Department of Communications and Electronics
Faculty of Engineering, Alexandria University

**Dr. Hanaa Abd-Elaziz Ebrahim**
Department of Switching
National Telecommunication Institute

Alexandria
2016

# Acknowledgement

This thesis with all its results would have never been achieved without ALLAH's will, and I am so grateful Allah for giving me the ability and strength to complete its final phases. I would also like to express my thanks for my parents for their effort and support throughout this thesis work.

First, I thank my advisor, **Prof. Mohamed Amr Mokhtar**, Faculty of Engineering, Alexandria University, for taking the time to discuss my research goals and providing me with helpful feedback. His guidance throughout the research and editorial help throughout the writing were invaluable.

Sincere acknowledgment and appreciation goes to my advisor, **Dr. Hanaa Abd El Aziz Ebrahim** Switching Department, National Telecommunication Institute and **Prof. Dr. Amr Ahmed El-Sherif,** Faculty of Engineering, Alexandria University, for their supervision, Knowledge, support, persistent encouragement, and for his support throughout the process of compiling this document.

I express my deepest gratitude to **Dr. Omayma Abd El Mohsen**, Head of Switching Department, National Telecommunication Institute, for being the resourceful initiator of this study; she provided the inspiration and advice necessary to create high quality research work.

I would also like to express my thanks to my work collages who were always supporting me and never hesitated a second to give me as much as they can from their knowledge, care, time and efforts.

I would also like to express my thanks to the Faculty of Engineering, Alexandria University and the National Telecommunication Institute for all the facilities that were available, and for the provided environment during my studies and research

Finally, with my love and gratitude, I want to dedicate this thesis to my family that support me ,especially to my brothers, my mother, and my father for their unconditional love , encouragement and support .

# Declaration

I declare that no part of the work referred to in this thesis has been submitted in support of an application for another degree or qualification from this or any other University or Institution.

# Abstract

The frequency spectrum is an essential resource of wireless communication. Special sections of the spectrum are used for military purposes, governments sell some frequency bands to broadcasting and mobile communications companies for commercial use, others such as ISM (Industrial, Science and Medical) bands are available for the public free of charge. As the spectrum becomes overcrowded, there seem to be two possible solutions: pushing the frequency limits higher to frequencies of 60 GHz and above, or reaggregating the densely used licensed frequency bands. The new Cognitive Radio (CR) approach comes with the feasible solution to spectrum scarcity. Secondary utilization of a licensed spectrum band can enhance the spectrum usage and introduce a reliable solution to its dearth. In such a cognitive radio network, secondary users can access the spectrum under the constraint that a minimum quality of service is guaranteed for the licensed primary users.

In this thesis, we focus on spectrum sharing techniques in cognitive radio network where there is a number of secondary users sharing unoccupied spectrum holes. More specifically, we introduce two collaborative cognitive radio networks in which the secondary user cooperate with the primary user to deliver the primary's data.

Firstly, we point out cognitive radio system with energy constraint imposed on the primary and secondary users. In this model, we study the cooperative cognitive radio network with energy harvesting primary and secondary users. The stable throughput region of the system is characterized for different Primary User (PU) and Secondary User (SU) energy harvesting rates. Moreover, the energy constrained cooperative system is compared with the cooperative system without energy constraints, and the non-cooperative energy constrained system. Additionally, we formulated mathematically the conditions at which the energy constrained system acts identically as the system with reliable power supply. Finally, we characterize the conditions for the system to switch between cooperative and non-cooperative modes.

Secondly, we propose a new method for cooperation and underlay mode selection in cognitive radio networks. We assume another system in which we study the spectrum sharing in the hybrid mode. In this model, we assume that the cognitive user could occupy the channel either in the overlay mode with cooperation, or with a small power in the underlay mode. The secondary user cooperates with the primary user in order to exploit the feedback message from the destination about the channel quality and getting more opportunities to access the channel. The acceptance of PU packets at the secondary user is regulated via admission control. This admission is function of the channel quality of the direct link between primary source and destination. To illustrate, the SU decision to cooperate with PU or not is depending on the PU channel quality. We formulate the problem as a belief/reward function and mathematically express these, belief and reward functions, seeking best performance of the system. Performance analysis displays that our proposed algorithm is better than conventional non-cooperative and cooperative systems in terms of secondary user throughput.

# Table of Contents

# List of Figures

# List of Figures

# List of Figures

# List of Tables

# List of Abbreviations

| | |
|---|---|
| ACK / NACK | Acknowledgment/ Negative- Acknowledgment |
| AWGN | Additive White Gaussian Noise |
| AC | Admission Control |
| BS | Base Station |
| BPU | Baseband Processing Unit |
| CSB | Cluster Supervision Block |
| CR | Cognitive Radio |
| DF | Decode-and-Forward |
| DPC | Dirty Party Coding |
| FA | False Alarm |
| FCC | Federal Communications Commission |
| FSMC | Finite State Markov Chain |
| HNC-MAC | Hybrid Non-Cooperative MAC |
| ISM | Industrial, Science and Medical |
| LO | Local Oscillator |
| MAC | Medium Access Control |
| MD | Missed Detection |

# List of Abbreviations

| | |
|---|---|
| MGF | Moment Generating Function |
| MPR | Multi-Packet Reception |
| MISO | Multiple-Input Single-Output |
| NTRA | National Telecommunication Regulation Authority |
| OSA | Opportunistic Spectrum Access |
| OCB | Optimal Cooperation Protocol |
| POMDP | Partially Observable Markov Decision Process |
| PU | Primary User |
| PDF | Probability Density Function |
| QoS | Quality of Service |
| RF | Radio-Frequency |
| SU | Secondary User |
| SDR | Software-Defined Radio |
| SS | Spread Spectrum |
| TDMA | Time Division Multiple Access |

# List of Symbols

| | |
|---|---|
| $\lambda_p$ | Arrival process at $Q_p$ |
| $\lambda_s$ | Arrival process at $Q_s$ |
| $\lambda_{ps}$ | Arrival process at $Q_{ps}$ |
| $\lambda_{ep}$ | Arrival process at $Q_{ep}$ |
| $\lambda_{es}$ | Arrival process at $Q_{es}$ |
| $\mu_p$ | Service rate of $Q_p$ |
| $\mu_s$ | Service rate of $Q_s$ |
| $\mu_{ps}$ | Service rate of $Q_{ps}$ |
| $a$ | service probability |
| $K$ | Secondary user cluster size |
| $\Gamma_m$ | SNR of level $m$ |
| $\pi_m$ | Steady state distribution of level $m$ |
| $\Lambda(m)$ | Level crossing rate of level $m$ |
| $M$ | Number of discrete channel quality levels |

# CHAPTER ONE
# INTRODUCTION

## 1.1 BACKGROUND

Recently, the extensive use of wireless communications collides with the shortage of resources required to establish communications. Spectrum scarcity coupled with the under-utilization of the licensed spectrum triggered the introduction of the concept of cognitive radio. Secondary utilization of a licensed spectrum band can enhance the spectrum usage and introduce a reliable solution to its dearth. Secondary users can access the spectrum under the constraint that a minimum quality of service is guaranteed for the primary users.

Recently, cooperation between the SU and PU has gained a lot of attention in cognitive radio research. Specifically, SUs act as relays for the PU's data while also trying to transmit their own data. The advantages of the cognitive transmitter acting as a "transparent relay" for the PU transmission are verified in terms of the system stability region enhancement, since the maximum allowed data arrival rates of both users are increased. Appropriate relaying improves the throughput of the primary user and can increase the transmission opportunities for the secondary user. Based on different multi-access protocols, cooperation schemes involved relaying either between the primary user and a secondary user or between secondary users themselves.

Investigating the impact of this protocol-level cooperation idea in wireless multiple-access networks is a challenge. The Cooperation works as follows: a packet is delivered to the primary destination either through a direct link or through cooperative relaying by intermediate secondary user. The secondary user has a statistically better channel to the destination. This introduces better spatial diversity to the primary users. The performance metrics of such a cooperative system include the stability region and the delay encountered by the primary and secondary data.

Energy harvesting (also known as power harvesting or energy scavenging) is the process by which energy is derived from external sources (e.g. solar power, thermal energy, wind energy, salinity gradients, and kinetic energy), captured, and stored for small, wireless autonomous devices, like those used in wearable electronics and wireless sensor networks. Energy harvesters provide a very small amount of power for low-energy electronics. While the input fuel to some large-scale generation costs money (oil, coal, etc.), the energy source for energy harvesters is present as ambient background and is free. For example, temperature gradients exist from the operation of a combustion engine. In urban areas, there is a large amount of electromagnetic energy in the environment because of radio and television broadcasting.

Energy harvesting from the working environment has received increasing attention in the research in wireless networks. Recent developments in this area could be used to replenish the power supply of wireless nodes. However, power management is still a crucial issue for such networks due to the uncertainty of stochastic replenishment.

Wireless networks can be self-sustaining by harvesting energy from ambient Radio Frequency (RF) signals. Recently, researchers have made progress on designing efficient circuits and devices for the RF energy harvesting. Many of research works related to the energy harvesting technology focused on performance analysis of previously studied system under the energy and power constraints [1-2]. Many questions appeared in the early studies of the energy harvesting systems. How energy harvesting and rechargeable batteries could be modeled, how the energy harvesting system acts to save the energy consumption rate, and do the energy-constrained systems perform like the unconstrained systems. Additionally, the optimal transmission power and network throughput under the new energy limitation had gained a lot of attention. In cooperative cognitive radio networks with energy harvesting, secondary users with better channel properties and enough energy harvesting rates provide a good choice for the primary users to get better service of their packets.

## 1.2 PROBLEM STATEMENT

A key challenge in operating cognitive radio in ad-hoc networks is how to allocate transmission power and spectrum access among the cognitive users. The first problem in this thesis concerns about the spectrum access and power challenges in cognitive radio networks. As for the spectrum access challenge, the cooperation between primary and secondary users was found as a solution for primary fading channel to improve the network's performance and attain better spectrum utilization. The secondary users cooperate with the primary users for their data transmission, getting mutual benefits for both users, such that, the primary user exploits the relaying capability provided by the secondary user, and the secondary user gets more opportunities to access the idle channel. As for the power challenge, primary and secondary users are supposed to be energy-constrained transceivers. These nodes do not have a reliable power supply for their data transmission and reception. Instead, they must have an alternative source of power, which is renewable and easy for implementation.

Energy harvesting is considered as a promising solution to alleviate such issues, which had received extensive attentions. Each energy-harvesting node has to have a device to harvest energy from the surrounding and save this energy for data transmission. The stored energy amounts are dissipated in the consecutive timeslots in channel sensing, packet transmission, and packet reception.

In the second problem, the target is to design a hybrid spectrum sharing method in cognitive radio, in which we are able to get the benefits from cooperation and underlay spectrum access. In this work, we propose a new hybrid access technique, which is switching between overlay with cooperation and underlay access modes to get better system performance by changing the access mode depending on the primary channel.

## 1.3 THESIS OBJECTIVES

In this work, we introduce two system models of cooperative cognitive radio networks. Firstly, we study a cooperative network with energy harvesting primary and secondary users. We characterize the stable throughput region of the system for different primary and secondary users' energy harvesting rates. In addition, we study in elaborate detail the impact on the system performance imposed by energy constraints.

Secondly, we propose a cooperative cognitive radio network in which the SU can access the channel in a hybrid way. The SU could cooperate with the PU or access the channel concurrently in the underlay mode, depending on the channel state of the primary link. The degree of cooperation between the two users is regulated via the Admission Control (AC) at the secondary user. The admission is function of the channel quality of the link between the primary source and destination. The behavior of the system exploits times when the primary channel is in good transmission conditions to allow the secondary user to access the channel concurrently with the PU with small transmission power.

## 1.4. THESIS OUTLINE

This thesis is organized as follows:

### 1.4.1 Chapter 2

In this chapter, we introduce an overview of cognitive radio networks, spectrum sharing, and energy harvesting. We begin with a survey about cognitive radio networks and related research areas. Then, we focus on spectrum sharing types and the corresponding literature. Finally, we illustrate in elaborate detail the concept of energy harvesting and its related work in cognitive radio networks.

### 1.4.2 Chapter 3:

This chapter investigates the maximum stable throughput of a cooperative cognitive radio system with energy harvesting primary and secondary users. The SU cooperates with the PU for its data transmission, getting mutual benefits for both users, such that, the PU exploits the SU power to relay a fraction of its undelivered packets, and the SU gets more opportunities to access idle timeslots. To characterize the system's performance, we compare the stable throughput regions at different energy harvesting rates.

Moreover, the energy constrained cooperative system is compared with the cooperative system without energy constraints, and the non-cooperative energy constrained system. Furthermore, we characterize the conditions for the system to switch between cooperative and non-cooperative modes. To simplify the analysis, a dominant system approach is used to obtain a closed form expressions for the system's stable throughput region.

### 1.4.3 Chapter 4:

This chapter illustrates cooperation and underlay mode selection in a cognitive radio network composed of single primary and secondary transceivers. We assume a cooperative system with admission control at the SU relay queue. The SU operation is hybrid between two modes; underlay and cooperation modes. The SU either cooperates with the PU to relay its data, or access the channel concurrently with the PU in the underlay mode. We assume the operation mode selection is based on the current state of the PU channel to its destination. Literature survey and related works are discussed in elaborate detail. We modeled the system links using FSMC method that matches the proposed admission control dependence on the channel quality. The system model is shown and the mathematical analysis is presented. The performance analysis and simulation show the improvement in network performance.

### 1.4.4 Chapter 5:

The conclusion of the proposed cooperative techniques and the future work of this thesis are discussed in this chapter.

# CHAPTER TWO
# COGNITIVE RADIO NETWORKS: SPECTRUM-SHARING APPROACH

## 2.1 INTRODUCTION

Most wireless networks are characterized by the static spectrum allocation technique, where a governmental agency (e.g. NTRA in Egypt) assigns the spectrum to licensed users on a long-term basis for large geographical areas. Because of the noticeable increase in spectrum demand by users, this policy is suffering from spectrum scarcity in certain bands. On the other hand, a large portion of this assigned spectrum is used sporadically (or underutilized), results in underutilization of a reasonable amount of the spectrum band [3]. Hence, the dynamic spectrum access approaches were proposed to produce a strong solution for the spectrum inefficiency problem.

The key technology of dynamic spectrum access approaches is the cognitive radio technology, which affords the capability to share the wireless channels with the licensed users in an opportunistic way. Cognitive radio networks are assumed to provide high bandwidth to users via dynamic spectrum access techniques. Cognitive radio networks, however, face severe challenges due to the high fluctuation in the spectrum availability, as well as the diversity in Quality of Service (QoS) requirements of various applications [4]. Since applications could be different in their requirements, real-time applications and error sensitive applications, different QoS levels are needed. In order to overcome these challenges, each cognitive radio user must be able to:

- Determine which portions are available of the spectrum of interest.
- Select the best available channel to produce the highest available rate.
- Coordinate the channel access with other cognitive radio users
- Vacate the used channel when a licensed user is sensed to be active [5]

These mentioned capabilities could be realized by the spectrum management function in cognitive radio nodes, which addresses the four main challenges:

- Spectrum sensing.
- Spectrum decision.
- Spectrum sharing.
- Spectrum mobility.

The following sections present the definition and the current research area of the spectrum management in cognitive radio networks. More specifically, we are concentrating on the discussion of the development of cognitive radio networks that require no modification in existing networks (licensed networks).

## 2.2 COGNITIVE RADIO TECHNOLOGY

Cognitive radio techniques provide the capability to share the spectrum in an opportunistic manner with the licensed networks. Therefore, cognitive radio is defined as a radio system that can change its transmitter and receiver parameters based on interaction with the current conditions of environment [3]. From the definition, two main aspects of the cognitive radio can be noticed [6]:

- Cognitive Capability: through the interaction with the radio environment in real-time manner, the portions of the spectrum that are free at a certain time or physical location can be identified. A cognitive radio user is able to use the temporally unused spectrum bands, which referred as spectrum hole or white space Fig. 2.1a.
- Reconfigurability: a cognitive radio node could be adjusted to transmit and receive on different ranges of frequencies, and use different access techniques powered by the hardware design [7]. By the aid of this capability, the best available spectrum band and the most appropriate operating channel parameters can be selected and reconfigured to use. The best available spectrum band is a design perspective, the best in terms of a cognitive radio user's design target.

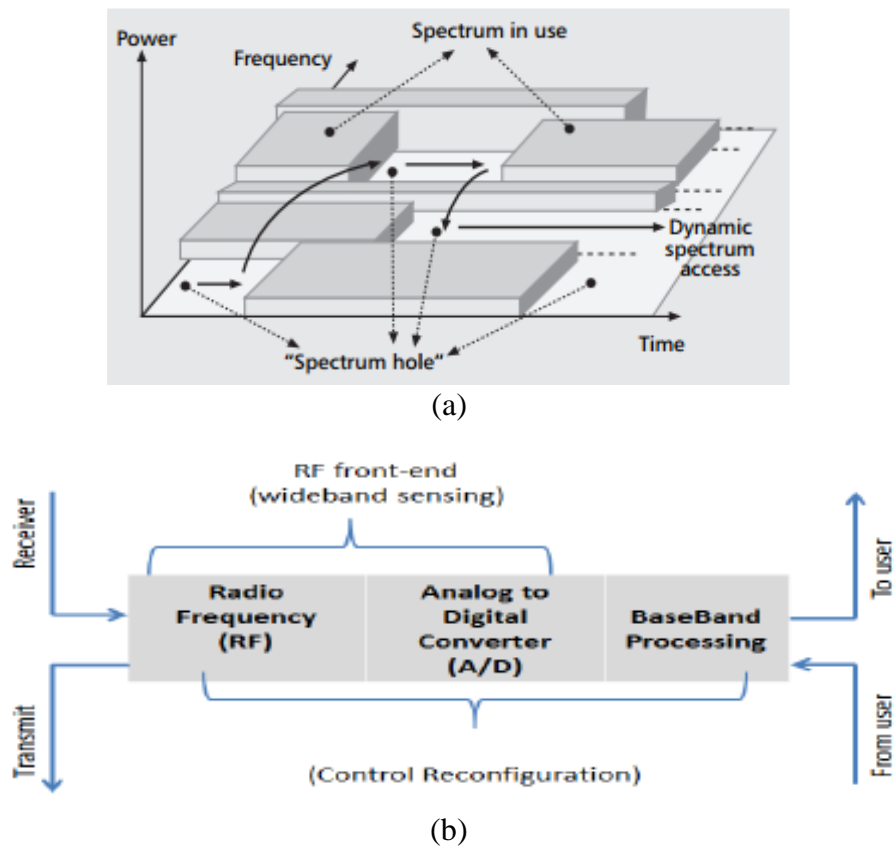

**Figure 2.1  Cognitive radio [3], a) the concept of spectrum hole
b) architecture of cognitive radio transceiver.**

In order to support these mentioned capabilities, a CR node requires a novel RF transmitter/receiver design. The main components of a cognitive radio user are the radio air interface and the Baseband Processing Unit (BPU) that were originally proposed for the Software-Defined Radio (SDR) Fig. 2.1b. In the RF front-end part, the received signal is amplified and digitized. In the BPU, the signal is modulated or demodulated. Each component must be adaptively configured to follow the time-varying RF environment. The novel characteristic of the cognitive radio transceiver node is the wideband RF front-end part that is able to concurrently sense multiple channels over a wide frequency range.

This functionality is mainly related to the RF hardware technology and limitations imposed on the industry, such as wideband antenna, power amplifier, and adaptive filter design either FIR or IIR filters. The RF hardware for the cognitive radio should be capable of being adjusted to any part of a large range of spectrum of the PU. Because the cognitive radio nodes receive signals from different transmitters operating at different SNR levels, bandwidths, and geometric locations. The challenge in cognitive radio transceiver design is that the RF front-end should have the capability to sense a relatively weak signal in a large dynamic range spectrum [8].

## 2.3 ARCHITECTURE OF COGNITIVE RADIO NETWORK

A complete description of the cognitive radio network architecture is mandatory for the development of protocols used to characterize dynamic spectrum allocation technology. The architecture of the cognitive radio system, as shown in Fig. 2.2, could be divided into two networks. These networks are the primary network and the secondary network.

The primary network is supposed to be originally existing network, where PUs have a license to operate in a certain frequency band. If primary network has a centralized infrastructure deployment, primary user activities are controlled through the centralized primary base station. Due to their higher priority in spectrum access, operations of secondary user should not affect the QoS targeted by the primary user.

The secondary users do not have a license to operate in a desired frequency band. Therefore, additional functions are imposed for the secondary users to share the licensed spectrum band concurrently with PUs. Secondary user network also could be equipped with SU base stations that provide single-hop connection to SUs. Additionally, cognitive radio networks may include spectrum brokers that play an important role in distributing the network resources among the different cognitive radio requirements [9].

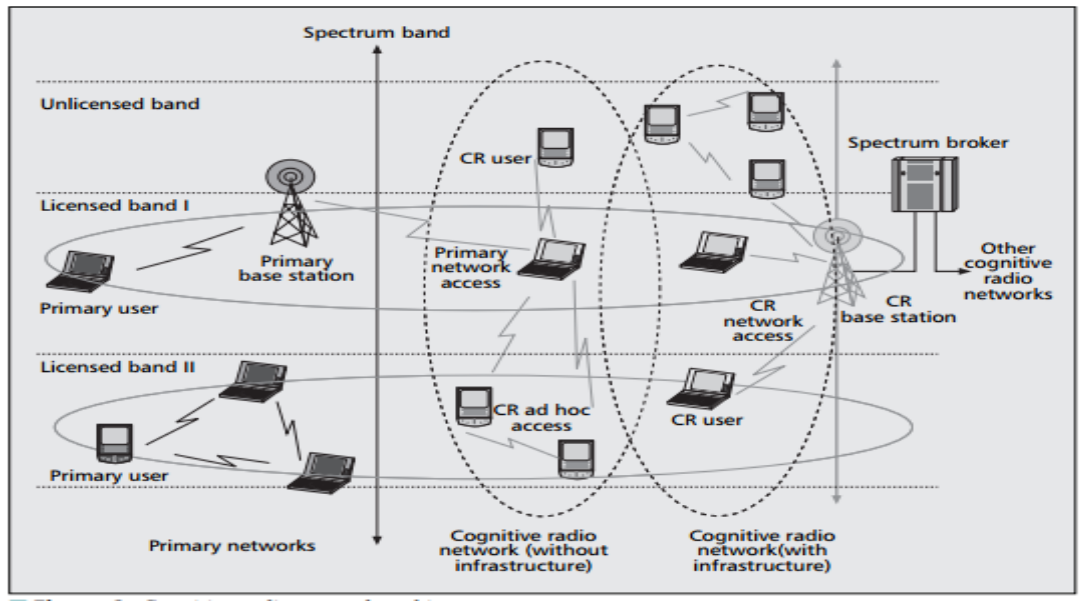

**Figure 2.2    Cognitive radio network architecture [3]**

## 2.3.1 Spectrum Heterogeneity

Secondary users are able to access both the licensed spectrum owned by the primary users and the unlicensed portions of the spectrum, which are used by contention through Opportunistic Spectrum Access (OSA) technology [3]. Therefore, the types of operation for cognitive radio networks could be divided to the licensed band operation and unlicensed band operation.

- Licensed band operation: the primary network owns this band in a long-term basis for large geographical areas. Here, cognitive radio users are focusing mainly on the detection of primary users' activities. The maximum transmission rate depends on the allowed transmission power, which in turn depends on the interference at nearby primary users' destinations. Primary users' abrupt appearance in the licensed band currently occupied by secondary users implies that secondary users should vacate that spectrum and move to next available spectrum immediately (spectrum mobility).
- Unlicensed band operation: all users have equal right to access the unlicensed spectrum (e.g. Wi-Fi spectrum use). Therefore, a sophisticated spectrum sharing techniques are required as a reliable contention method for cognitive radio users to compete for the unlicensed band.

### 2.3.2 Network Heterogeneity

As shown in Fig. 2.2, the cognitive radio users have the opportunity to access the idle spectrum (either licensed or unlicensed) in three different ways [3]:

- Cognitive radio network access: cognitive radio users can access the cognitive radio base station, on both licensed and unlicensed spectrum bands. Since all interactions occur in the secondary network, the SUs' spectrum sharing technique could be independent of the technique of the primary network.
- Cognitive radio ad hoc access: In this access method, SUs can communicate with each other through an ad hoc connection manner on both the licensed and unlicensed bands.
- Primary network access: In this access method, SUs can also access the PUs' base station through the licensed spectrum band. Unlike for first two access methods, SUs require a Medium Access Control (MAC) protocol, to support roaming over multiple PU networks with different access techniques.

As shown in the cognitive radio architecture Fig. 2.2, various functionalities are required to support spectrum management in cognitive radio networks.

### 2.3.3 Spectrum Management Framework

Cognitive radio networks impose many challenges due to their coexistence with the primary network as well as diverse QoS requirements. Thus, new spectrum management functions are required for the CR networks with the following design challenges [3]:

- Interference avoidance: cognitive radio networks should avoid interference with primary users.
- Quality of service awareness: to decide on what is the appropriate spectrum band to be occupied, cognitive radio networks should support QoS-aware concept, considering the dynamic and heterogeneous spectrum conditions.

- Seamless communication: cognitive radio networks should also provide seamless communication irrespective of the appearance of primary users.

To overcome these challenges, different functionalities required for spectrum management in CR networks [10]. Spectrum management process consists of four major parts:

- Spectrum sensing: a cognitive radio user can allocate only an unused portion of the spectrum. Therefore, a cognitive radio user should monitor the available spectrum, capture their information, and then detect spectrum holes.
- Spectrum decision: based on the spectrum availability, cognitive radio users can allocate a channel. This allocation not only depends on spectrum availability, but also is determined based on internal (and possibly external) policies.
- Spectrum sharing: because there may be multiple cognitive radio users trying to access the spectrum, cognitive radio network access should be coordinated to prevent multiple users colliding in overlapping portions of the spectrum.
- Spectrum mobility: cognitive radio users are regarded as visitors to the spectrum. Hence, if a primary user requires the specific portion of the spectrum in use, the communication must be continued in another vacant portion of the spectrum.

The spectrum management framework for cognitive radio network communication is illustrated in Fig. 2.3. It is evident from the significant number of interactions that the spectrum management functions require a cross-layer design approach. In the following sections, we discuss the four main spectrum management functions.

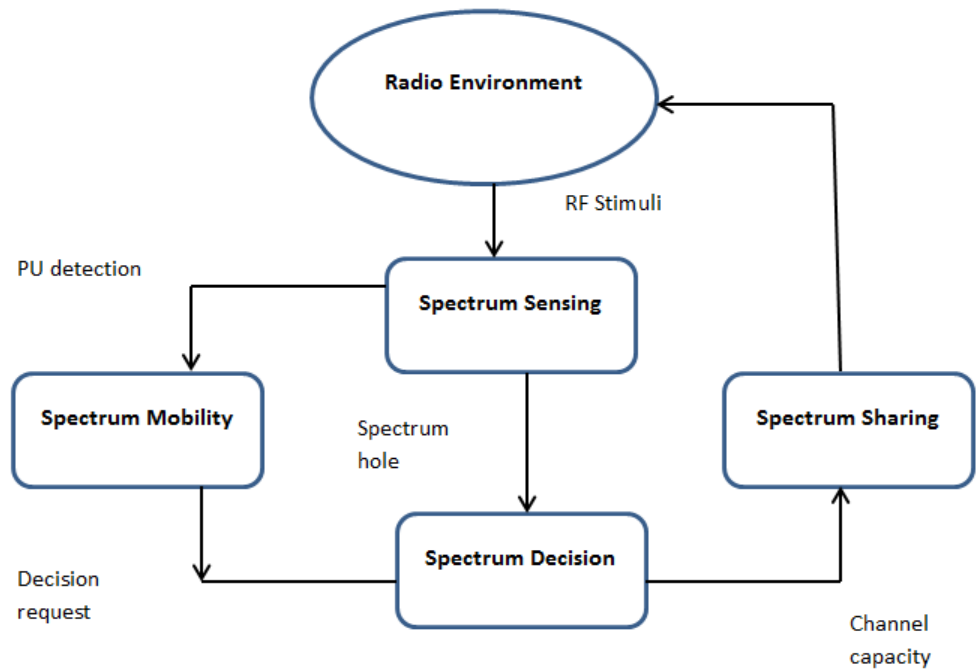

**Figure 2.3    Spectrum management framework [10]**

## 2.4 SPECTRUM SENSING

A cognitive radio user is designed to be aware of changes in its surroundings environment, this makes spectrum sensing an important requirement for the feasibility of cognitive radio networks. Spectrum sensing enables cognitive radio users to adapt dynamically to the environment and activity changes by detecting spectrum holes seeking not to cause interference to primary users. This can be achieved through a real-time wideband sensing capability in the cognitive radio transmitter to detect the primary signals in a wide spectrum range. In general, spectrum sensing techniques can be classified into three groups [11]: primary transmitter detection, primary receiver detection, and interference temperature management.

## 2.5 SPECTRUM DECISION

Cognitive Radio networks require the capability to decide which spectrum band to occupy among the available frequency bands according to the QoS requirements. This is why the notion here called spectrum decision and constitutes an important topic in cognitive radio network. Spectrum decision is related to the channel characteristics and operations of primary user's activities. Spectrum decision usually consists of two steps: first, each spectrum band is characterized, based on not only local observations of CR users but also statistical information of primary networks. Then, based on this characterization, the most appropriate spectrum band could be chosen.

## 2.6 SPECTRUM MOBILITY

After a cognitive radio user captures the best available spectrum, primary user activity on the selected spectrum may necessitate that the SU changes its operating spectrum band(s), which is referred to as spectrum mobility. Spectrum mobility gives rise to a new type of handoff in CR networks, spectrum handoff. Protocols for different layers of the network stack must adapt to the channel parameters of the operating frequency. Moreover, they should be transparent to spectrum handoff and the associated latency.

Each time a CR user changes its frequency of operation, network protocols may require modifications to the operation parameters [14]. The purpose of the spectrum mobility management in CR networks is to ensure smooth and fast transition leading to minimum performance degradation during a spectrum handoff. An important requirement of mobility management protocols is the information about the duration of a spectrum handoff. The sensing algorithm can provide this information. After the latency information is available, the ongoing communications can be preserved with only minimum performance degradation.

Although the mobility-based handoff mechanisms that have been investigated in cellular networks may lay the groundwork in this area, there are still open research topics to be investigated. In order to improve the spectrum utilization of the licensed spectrum, the cognitive radio is supposed to dynamically exploit the opportunistic primary frequency spectrum and provide an efficient spectrum handoff technique. An efficient spectrum handoff technique has been suggested in [14] based on the Partially Observable Markov Decision Process (POMDP) to estimate the network information by partially sensing the frequency spectrum. This POMDP-based Spectrum Handoff (POSH) scheme is proposed to determine the optimal target channel for spectrum handoff according to the observable channel state information. By adopting the policy resulted from the POSH algorithm for target channel selection, the SU could switch to a PU frequency with the lowest expected waiting time.

## 2.7 SPECTRUM SHARING

The broadcast nature of the wireless channel requires the coordination of transmission attempts between the cognitive radio users. Correspondingly, spectrum sharing should include much of the functionality of a MAC protocol. Spectrum sharing can be classified based on four classification criteria: the architecture, spectrum sharing inside a CR network or among multiple coexisting CR networks, spectrum allocation behavior, and finally spectrum access technique that is the scope of this thesis.

### 2.7.1 Spectrum Sharing Architecture

The first classification is based on the architecture, which can be centralized or distributed:

- Centralized spectrum sharing: the spectrum allocation and access procedure are controlled by a central entity. Moreover, measurements of the spectrum allocation at the distributed nodes can be forwarded to the central entity, and a spectrum allocation map is established. The central entity can lease spectrum to users in a limited geographical region for a specific amount of time [8].
- Distributed spectrum sharing: spectrum allocation is based on local policies that are performed by each secondary user in a distributed manner [14]. Distributed solutions also are used between different networks such that a base station competes with its interferer BSs according to the QoS requirements of its users. The comparison between centralized and distributed solutions reveals that distributed solutions closely follow the centralized solutions in the performance, but at the cost of message exchanges between nodes [15].

### 2.7.2 Intranetwork and Internetwork Spectrum Sharing

The second classification is based on does the spectrum sharing is inside a CR network (intranetwork spectrum sharing) or among multiple coexisting CR networks (internetwork spectrum sharing) [3]:

- Intranetwork spectrum sharing: these techniques focus on spectrum allocation between entities of a cognitive radio network, as shown in Fig. 2.4.

Accordingly, users of a cognitive radio network try to access the available spectrum without causing interference to the primary users.

- Internetwork spectrum sharing: the cognitive radio architecture enables multiple systems to be deployed in overlapping locations and spectrum, as shown in Fig. 2.4.

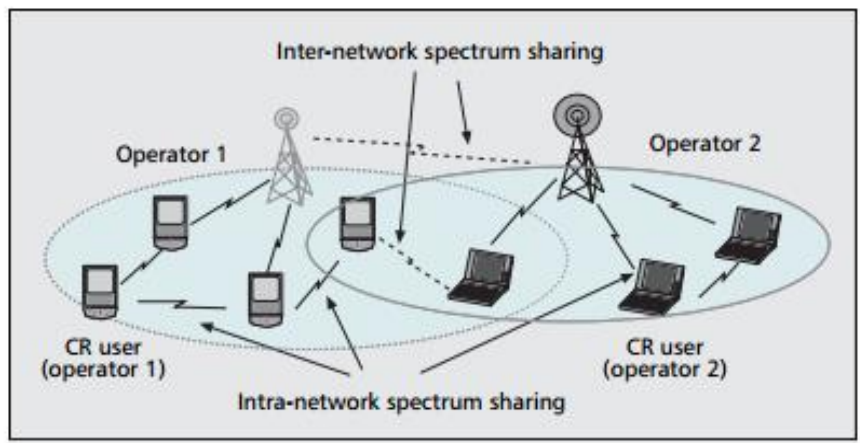

**Figure 2.4　　Channel structure of the multi-spectrum decision [3].**

## 2.7.3 Spectrum Sharing Allocation Behavior

Spectrum access can be cooperative or non-cooperative. Cooperative spectrum sharing (collaborative) exploits the interference measurements of each node such that the effect of the communication of one node on other nodes is considered. In non-cooperative spectrum sharing, only a single node is considered [16]. Because interference in other CR nodes is not considered, non-cooperative solutions may result in reduced spectrum utilization. However, these solutions do not require frequent message exchanges between neighbors as in cooperative solutions. Cooperative approaches generally outperform non-cooperative approaches, as well as closely approximating the optimum spectrum utilization [15].

## 2.7.4 Spectrum Sharing Access Technology

The fourth classification is based on the access technology [17]:

- Overlay spectrum sharing: secondary users access the network using a portion of the spectrum that has not been occupied by the licensed users. This minimizes interference to the primary users.
- Underlay spectrum sharing: Spread Spectrum (SS) techniques are exploited such that primary users regard the transmission of CR nodes as noise. Underlay techniques can provide more access opportunities at the cost of a slight increase in complexity and interference. The realization of efficient and seamless spectrum sharing technique in cognitive radio networks still has many open research areas, such as common control channel [18] and dynamic radio range.

## 2.8 COOPERATIVE COGNITIVE RADIO:

As introduced in section 2.7, classification of spectrum sharing based on the access technology includes the overlay mode and the underlay mode (or hybrid between underlay and overlay). As an extension, the SU can cooperate with the PU to relay the PU data. Recently, cooperation between the SU and PU has gained a lot of attention in cognitive radio research [17-18]. Specifically, SUs act as relays for the PU data while also trying to transmit their own data. Cooperative relaying by intermediate nodes that have a statistically better channel to the destination is the intended way to enhance the PU delivery of data while increasing the opportunity for the SU to access idle channel.

In [19], advantages of the cognitive transmitter acting as a "transparent relay" for the PU transmission are investigated. Authors proved that the stability region of the system would increase in terms of the maximum allowed arrival rates of both the PU and SU. In addition, it is shown by the power allocation mechanism that the maximum allowed transmission power for the SU would increase. The secondary transmitter is allowed to forward packets of the primary user that it has been able to decode, but have not been successfully received by the primary destination. The benefit beyond this choice is that, if the propagation channel from the primary transmitter to the secondary transmitter is advantageous with respect to the direct channel; having its packets relayed by the secondary user source can help emptying the queue of the primary faster than the direct link and creating transmitting opportunities for the secondary.

A system model consisting of two types of nodes, primary and secondary nodes, is introduced. The main requirement for cooperation is that the activity of the secondary nodes should be "transparent" to the primary user, i.e. not to interfere with the primary user's activity. Secondary user can overhear feedback messages from PU destination to the PU source, Acknowledgment/ Negative-Acknowledgment (ACK/ NACK) messages. If a packet is successfully decoded by the SU while the PU destination could not receive it correctly, the secondary user sends independent ACK message to the primary source indicating that it would bear the responsibility to deliver this packet in the subsequent timeslots. System model of [19] is shown in Fig. 2.5.

Whenever the secondary node senses an idle slot, it transmits a packet from queue $Q_s$ with probability $\varepsilon$ and from the second queue $Q_{ps}$ with probability $(1-\varepsilon)$. Analysis of this system verified that, the maximum throughput of the secondary user increased in the cooperative system model with sustaining the primary user stability as in the non-cooperative system. In addition, the maximum secondary user transmission power is maintained for higher primary user arrival rates than in the non-cooperative system.

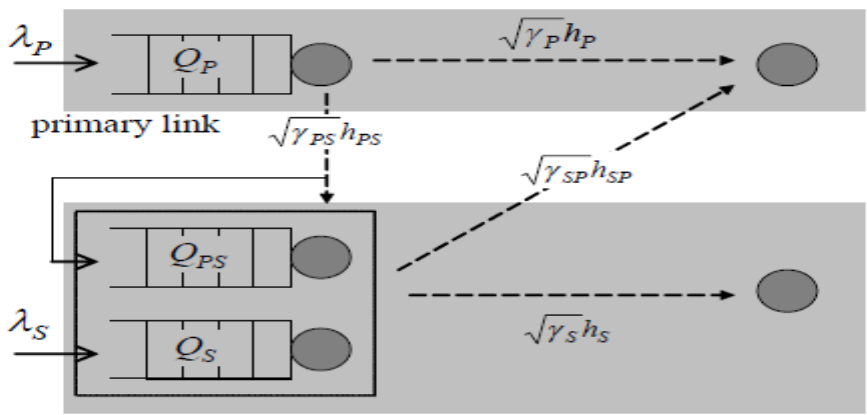

**Figure 2.5**   **The secondary transmitter may act as a relay for the primary link [19].**

The stability and throughput of a two-user cognitive radio system with multicast traffic is discussed in [20]. One of the users could be used as a relay for the packets of the other user that are not received at the destination through the first transmission attempt. It is shown that the stability and throughput regions of this cooperative system are wider than that of the non-cooperative system, which leads into an improvement for both users of the system.

A time-slotted, multiple-access system S, consisting of two nodes S1 and S2 and two destinations d1 and d2, is proposed. Authors allowed for the possibility of node S2 acting as a cooperative relay for node S1, such that if it receives the packet from S1 before both d1 and d2, it assumes the responsibility of ensuring that the packet is received at both destinations. In this system, the channel between S2 and the destinations is better than the one between S1 and the destinations. Because S2 can improve the chances of S1's data to be received, it was shown that the stability region of the cooperative system is outer bound to the non-cooperative system.

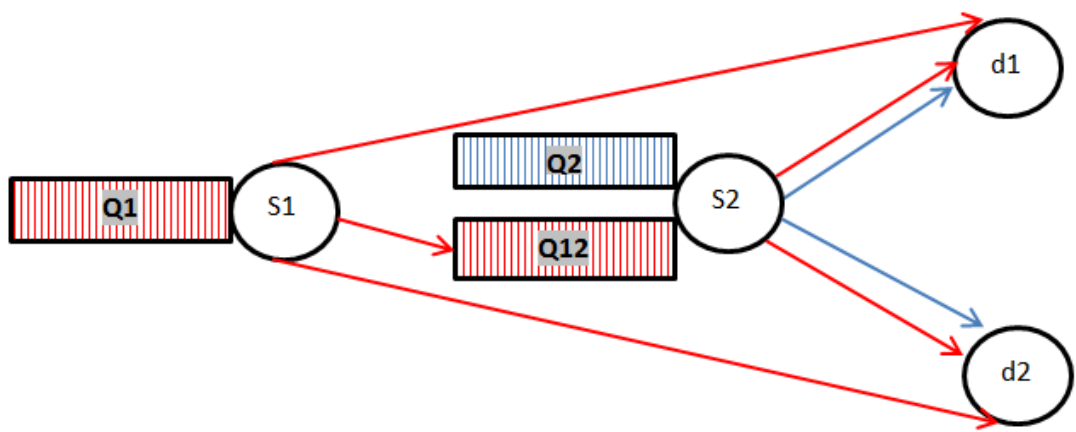

Figure 2.6     System model of [20]

The protocol design for cognitive cooperative system with multi secondary users is proposed in [21]. In contrast with previous cognitive configurations, the channel model assumed a cluster of secondary users that perform both sensing process for transmitting opportunities and relaying data for the primary user.

Authors designed and characterized protocol for cognitive radio with many secondary users. A new cognitive structure is introduced, in which a cluster of nodes sense the radio spectrum for transmission opportunities. The cognitive cluster is equipped with a common (for all the nodes of the cluster) relaying queue in order to relay incorrectly received data for the primary destination. System model of [21] is shown in Fig. 2.7.

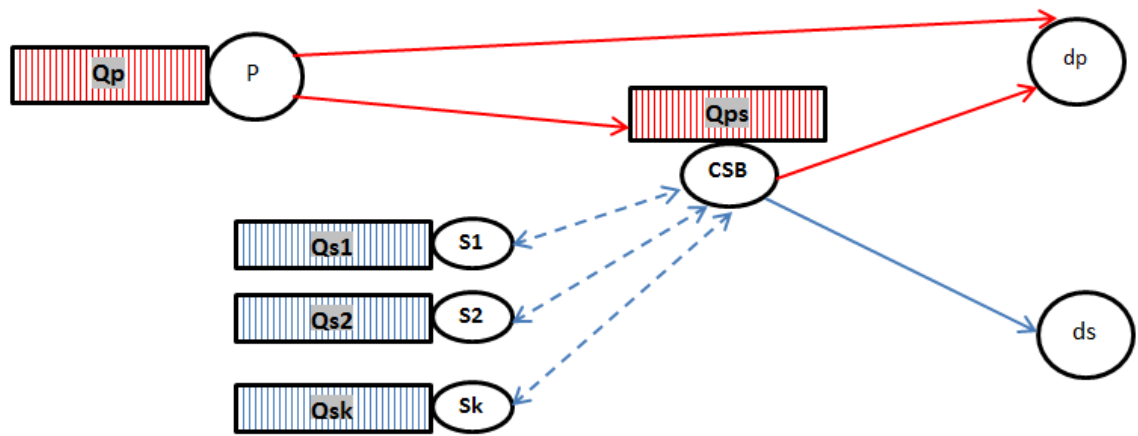

**Figure 2.7    System model of [21]**

The basic problem was to study the interplay among the primary user, the common relaying queue, and the secondary queues as well as the optimization target. Previously reported single-user schemes did not provide efficient solutions for the scenario of multiple secondary users system and motivate the exhibition of new cooperative protocols that take into account the multi-user nature. A cooperative scheme is proposed in which simultaneous transmissions is enabled for the primary user and a secondary user, or two secondary users. Both transmitters can transmit the same primary data by creating a virtual Multiple-Input Single-Output (MISO) system or a combination of primary and secondary data by using Dirty Party Coding (DPC) [22].

The cognitive cooperative system of multiple primary users and one relaying secondary user is discussed in [23]. Primary users transmit packets in orthogonal sub-channels. According to the cognitive principle, the secondary activity cannot interfere with the primary performance. The secondary user makes use of the spectrum only when sensed idle.

Based on the proposed MAC, authors derive the stable throughput and delay expressions of each primary user with secondary relaying. Results revealed that secondary relaying could increase the primary and secondary throughput and reduce the delay when designing an appropriate relaying probability. System model of [23] is shown in Fig. 2.8.

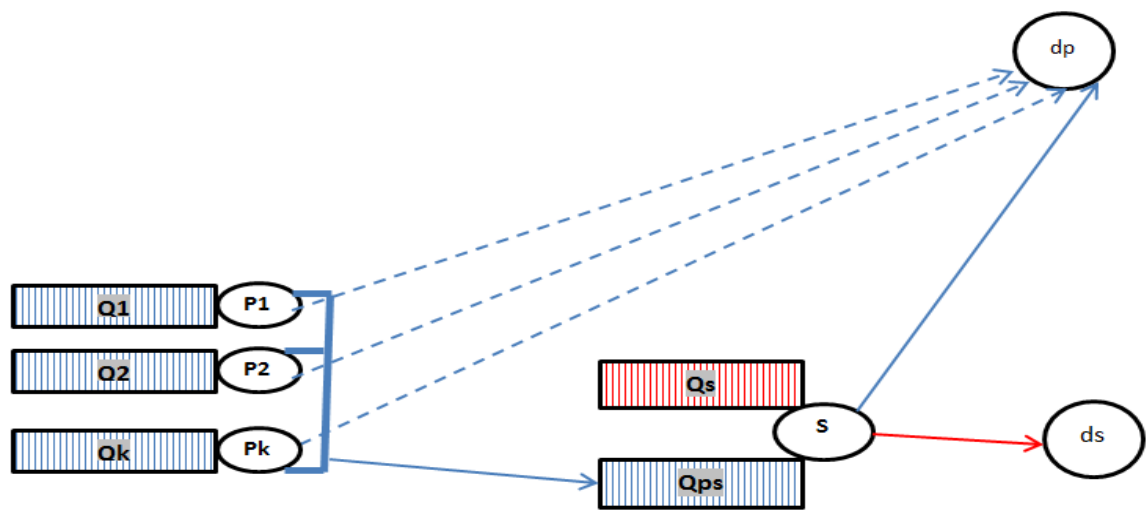

**Figure 2.8　　System model of [23]**

Fundamental throughput and delay tradeoff in cooperative cognitive radio system is studied in [24]. Authors focused on the class of randomized cooperative policies, whereby the secondary user serves either its own data queue or the relay queue of the primary user relayed data with certain service probability. Admission control is assumed at the relay queue whereby the PU's packet, which is received and decoded correctly at the relay queue, is admitted to the relay queue with certain admission probability ($Pa$). The proposed policy illustrates the fundamental tradeoff between the delay of the PU packets and the SU packets.

The complete stable throughput region of the non-work conserving system obtained via taking the union over all possible values of the admission probability. Cooperation is shown to extend the stability region area of the system. Moreover, it has been shown that increasing the serving probability from the SU own queue, $Pq$, is always in favor of the SU throughput and delay. This behavior is irrelevant to the choice of concurrent admission probability.

Authors suggest using $Pa$ as a switch with which it could be decided to cooperate or not to cooperate to optimize the performance of the PU depending on the value of $Pq$ and the channel qualities. System model of [24] is shown in Fig. 2.9.

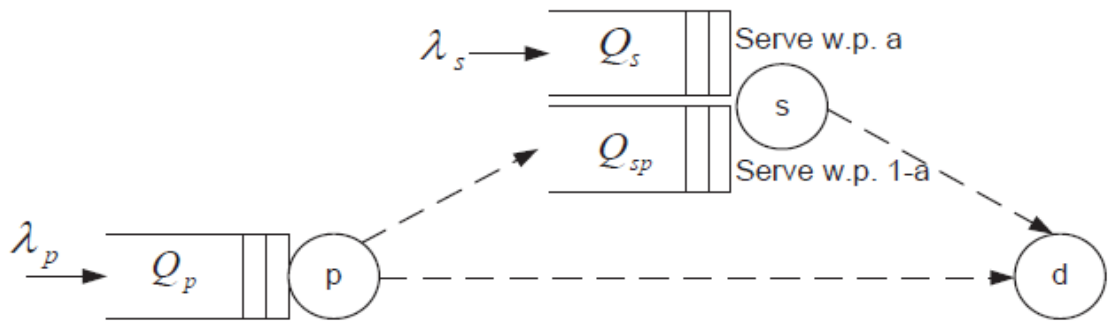

**Figure 2.9**　　System model of [24]

## 2.9 ENERGY HARVESTING

Energy harvesting, has gained significant interest in both academia and industry. Energy harvesting refers to the capture of ambient energy, its conversion into a useable form and its storage for future use. Various types of energy could be harvested e.g. mechanical energy, electromagnetic energy, and thermal energy [25]. Energy harvesting from radio frequency signals or electromagnetic energy could exploit the activity of primary users to harvest energy. Block diagram repressing energy-harvesting functions is plotted in Fig. 2.10.

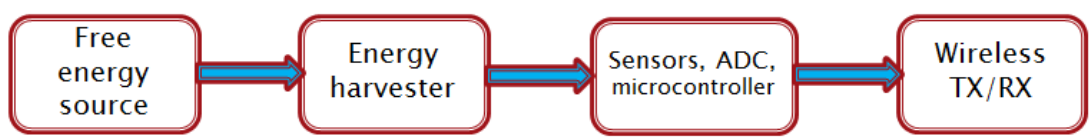

**Figure 2.10**　　Energy harvesting block diagram

Energy limitations and constraints on transmission power have recently gained a lot of interest. More specifically, energy harvesting appeared as alternate for the reliable power supply in cognitive radio system. Several articles have discussed energy harvesting as a solution for hard-wiring or replacing batteries of rechargeable wireless devices [26-28].

Non-cooperative energy harvesting cognitive radio network with multi-packet reception channel model is introduced in [29]. In that work, the maximum stable throughput region of the system is studied using stochastic dominance technique for optimizing the utilization of a cognitive shared channel. A cognitive protocol on the general multi-packet reception model of the channel is adopted where the primary transmission may succeed even in the presence of secondary transmission. As a result, the secondary user can increase its throughput through not only utilizing the idle periods of the primary user but also randomly accessing the channel by some probability. Additionally, authors analyzed two different models. First, the secondary transmitter harvests energy for transmission, and second, both primary user and secondary user are equipped with rechargeable batteries. Then the maximum stable throughput region for the two models is studied. System model of [29] is shown in Fig. 2.11.

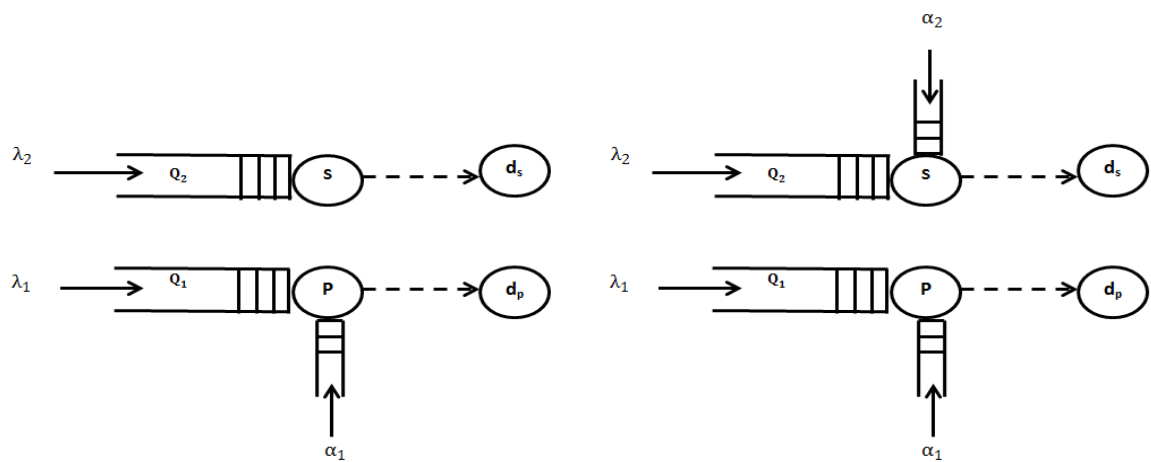

**Figure 2.11     System model of [29]**

In [30], network-layer cooperation in a wireless three-node network with energy harvesting nodes is modelled. Energy harvesting is presented as a buffer that stores harvested energy to serve a bursty data traffic queues. Since modelling energy harvesting in each node as a queue, authors studied the interaction between data and energy queues when arrival rates at the queues are known. The maximum allowable throughput of the source and the required transmission power for both a non-cooperative and Decode-and-Forward (DF) cooperative schemes are derived. It was proved that the cooperation achieves a higher maximum stable throughout than direct link for scenarios with poor energy arrival rates.

A cognitive radio network with one PU and one SU, with their transmitters operate in time-slotted mode, is discussed in [31]. The SU harvests energy from ambient radio signal and follows up a save-then-transmit protocol. In this scenario, the SU's optimal cooperation strategy was to choose the optimal decision, whether to cooperate or not. Spending how much time on energy harvesting and allocating how much power for relaying are calculated from the optimal action. System model of [31] is shown in Fig. 2.12.

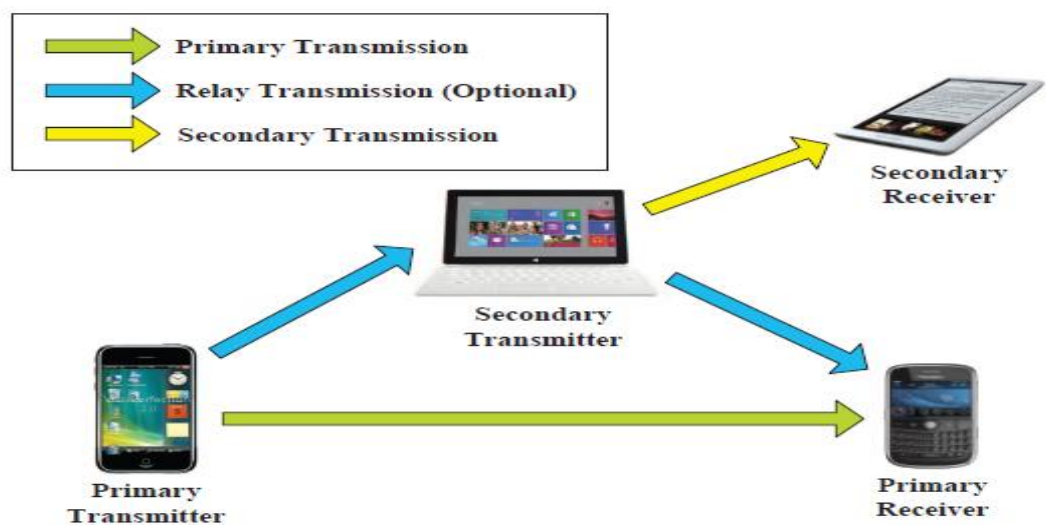

**Figure 2.12    System model of [31]**

In the non-cooperative mode, the SU first collects energy from ambient radio signal and is allowed to transmit its own data only when the PU's transmission is finished.

The time to save harvested energy is shown to be equal or greater than the time of PU's packet transmission. During the cooperative transmission, the PU transmitter transmits its data to the SU transmitter as well as the PU receiver. The SU source spends $tc/2$ for data decoding from the PU source, and consecutive $tc/2$ to forward the decode PU's data. Timeslot structure of non-cooperation mode and cooperation mode are shown in Fig. 2.13.

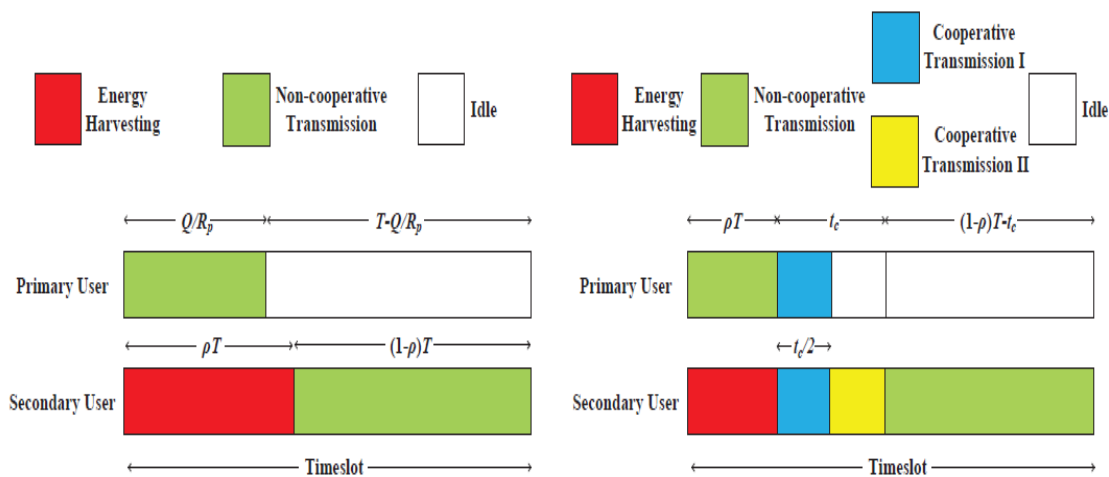

Figure 2.13   Timeslot structure of non-cooperation and cooperation modes [31]

Queues stability in a slotted ALOHA random access network with energy limitation was studied in [32]. Two nodes are equipped with a battery for energy storage. The stability region of this system is compared with the stability region of the system without energy constraints. It is assumed that nodes can exploit the feedback information for collision detection in next timeslots.

When the collision occurs, the destination stores the collided packets, and sends NACK to both sources. Consequently, after the NACK is heard, one of the nodes retransmits its collided packet while the other is silent. The common destination uses the retransmitted packet and the old stored collided packets to recover the two packets involved in the collision in two successive transmissions. Therefore, the two nodes are served in only two transmissions, even in case of collision, but the retransmitting node has used more energy in this process. Authors modelled each node as two queues, the first queue for storing packets and the second model the energy in the battery.

The stability region of this system is illustrated with the aid of the dominant system approach. The obtained stability region is compared with that of Time Division Multiple Access (TDMA) system and a random access network without energy constrained. The reduction in the stability region due to the finite energy is shown. This reduction is shown in three scenarios of the energy constrains.

The stability region of this system is illustrated with the aid of the dominant system approach. The obtained stability region is compared with that of TDMA system and a random access network without energy constrained, and the reduction in the stability region due to the finite energy is investigated. This reduction is shown in three scenarios of the energy constrains. First, energy constraint on the first source, second, energy constraint on the second source, and finally, energy constraints on both users are compared.

A cognitive radio network with energy harvesting secondary user is assumed in [33] approaching the cross layer design of MAC and physical layers. A cross layer design involved both random spectrum sensing and access protocol for the SU that exploits the primary link's feedback. Unlike previous works, authors did not assume perfect feedback. Instead, they take into account the more practical possibilities of overhearing unreliable feedback signals and impose spectrum-sensing errors. Interference-based channel model, where receivers are equipped with multi-packet reception capability, is assumed. Power allocation at the SU is performed with the objective of maximizing the secondary throughput under constraints that maintain PU required QoS.

A time-slotted cognitive radio setting with energy harvesting primary and secondary users is studied in [34]. At the beginning of each timeslot, the secondary user probabilistically chooses the spectrum sensing duration from a predefined set of sensing durations. If the primary user is sensed to be idle, the secondary user accesses the channel immediately. The CR user optimizes for the sensing duration probabilities in order to maximize its mean data service rate with constraints on the stability of the primary and cognitive queues.

# CHAPTER THREE
# COOPERATIVE COGNITIVE RADIO NETWORK WITH ENERGY HARVESTING: STABILITY ANALYSIS

## 3.1 INTRODUCTION

In this chapter, we investigate the maximum stable throughput of a cooperative cognitive radio system with energy harvesting primary and secondary users. It was shown in the literature that cooperative cognitive radio networks purvey better system performance than the non-cooperative networks. Since supply of reliable power sources is a challenge to provide in wireless nodes at certain circumstances, energy harvesting was found as the alternative solution for such operation. Energy harvesting was introduced as a hot research area in wireless communication and cognitive radio. In this chapter, we combine the cooperative spectrum sharing techniques with the energy harvesting capable primary and secondary users to attain a reliable wireless transmission in the absence of reliable power supplies. The secondary user cooperates with the primary user for its data transmission, getting mutual benefits for both users, such that, the primary user exploits the SU power to relay a fraction of its undelivered packets, and the secondary user gets more opportunities to access idle timeslots.

Results reveal that, the energy-constrained system (system with energy harvesting) identically acts as the non-energy-constrained system in certain conditions that are characterized. Conditions show that the energy harvesting may be provide sufficient energy for transmission the data of the proposed system at low to moderate arrival rates at both users. Consequently, the deficit of energy required for transmission appears at higher arrival rates at the primary and secondary user.

In order to investigate the impact of scaling the proposed system, we introduce a system composed of K secondary users. A cluster is composed of K secondary user with Cluster Supervision Block (CSB) to control the activity and transmission selection among the K SUs nodes. The secondary user cluster, with energy harvesting capability on all nodes, transmits data to a common destination. The existence of multi-SUs in the SU cluster introduces channel diversity to the primary user packets to be relayed.

## 3.2 SYSTEM MODEL

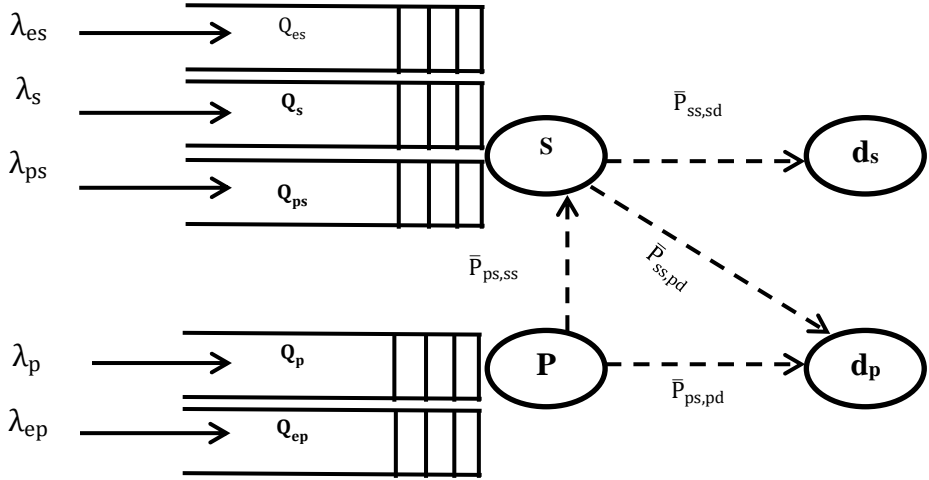

**Figure 3.1    System model**

Fig. 3.1 depicts the proposed system model under consideration. The system is composed of one PU and one SU, assumed to harvest energy from the environment. Primary user has two queues, $Q_p$ and $Q_{ep}$. $Q_p$ is an infinite capacity buffer for storing the PU's fixed length packets. The arrival process at $Q_p$ is modeled as Bernoulli arrival process with mean $\lambda_p$ [packets/slot]. $Q_{ep}$ models the PU's battery, assumed to have an infinite size to store the harvested energy. Energy is assumed to be harvested in a certain unit and one unit of energy is consumed in each transmission attempt, assuming one unit of energy is equal to the transmission power of the source multiplied by the time of packet transmission. The energy harvesting process is modeled as a Bernoulli arrival process with mean $\lambda_{ep}$. These processes are independent, stationary over timeslots.

Considering the SU, it is represented by three queues: $Q_s$, $Q_{ps}$, and $Q_{es}$. $Q_s$ is an infinite capacity buffer for storing the SU's own packets. The secondary relay queue, $Q_{ps}$, stores the PU's packets successfully received by the SU when the channel between the PU transmitter and receiver is in outage. $Q_{es}$ is the SU battery of infinite size storing the harvested energy. The arrival processes at the two queues, $Q_s$ and $Q_{es}$, are modeled as Bernoulli arrival process with means $\lambda_s$ and $\lambda_{es}$, respectively. Time is slotted, and a packet transmission takes one timeslot. Therefore, the average arrival rates $\lambda_p$ and $\lambda_s$ [packets/slot] lie in the interval [0, 1]. The arrival processes at each user are independent across successive timeslots (i.i.d). The average arrival rates $\lambda_{ep}$ and $\lambda_{es}$ [energy packets/slot] lie in the interval [0, 1].

The PU transmits a packet from $Q_p$ whenever it is non-empty. If the channel between the PU transmitter and receiver is not in outage, then the PU receiver successfully decodes the packet and the packet departs the system. It is assumed that the SU can overhear the ACK/NACK from the PU receiver. In the timeslots when the channel between the PU transmitter and receiver is in outage, if the SU received the PU packet correctly, the packet will be stored in the relay queue and the SU will bear the responsibility to deliver this packet. If the channel between the PU and SU also is in outage, PU will try to retransmit the packet in a subsequent timeslot. Any transmission or retransmission from $Q_p$ requires that $Q_{ep}$ be non-empty. The SU is assumed to perform perfect sensing. Whenever the channel is sensed to be idle, the secondary has two data queues to transmit a packet, specifically $Q_s$ and $Q_{ps}$. The SU is assumed to transmit a packet from $Q_s$ with probability $a$, or from $Q_{ps}$ with the complement probability $\bar{a} = 1 - a$.

### 3.2.1 Physical Channel

All wireless links exhibit fading and Additive White Gaussian Noise (AWGN). Outage of a link occurs when the instantaneous capacity of that link is lower than the transmitted spectral efficiency rate. Let $\overline{P}_{j,k}$ denote the probability that the channel is not in outage between j and k, where $j \in \{ps, ss\}$, $k \in \{pd, ss, sd\}$, also ps, ss, pd, and sd represent the PU source, SU source, PU destination, and SU destination, respectively. On contrary, $P_{j,k}$ denote the probability that the channel is in outage between j and k.

## 3.2.2 Energy Harvesting stable throughput regions

In this section, the stable throughput region of the system under consideration is characterized. This region is bounded by the maximum arrival rates at the PU and SU when the two queues, $Q_p$ and $Q_s$, are stable. The stability of the queue is identified by Loyne's theorem [35]. The theorem states that if the arrival and service processes of a queue are stationary, then the queue is stable if the arrival rate is strictly less than the service rate. For any queue in the system, the stability requires that:

$$\lambda_i < \mu_i, \tag{3.1}$$

where $i \in \{p, ps, s, ep, es\}$, and $\mu_i$ refers to the service rate of the $i^{th}$ queue.

Starting with the PU data queue stability, a packet is serviced from $Q_p$ if it is successfully decoded by the PU destination, or by the SU. We then have,

$$\mu_p = (\overline{P_{ps,pd}} + P_{ps,pd}\overline{P_{ps,ss}})\Pr\{Q_{ep} \neq 0\} \tag{3.2}$$

The probability that $Q_{ep}$ is empty, $P_{(Qep=0)}$, is obtained from the Little's law, [36], by $(1 - \lambda_{ep}/\mu_{ep})$, where $\lambda_{ep}$ and $\mu_{ep}$ denotes the arrival and service rate of $Q_{ep}$, respectively. It is obvious that the service rate of the PU battery queue $Q_{ep}$ depends on whether the PU data queue $Q_p$ is empty or not. Similarly, the service rate of $Q_p$ depends on the state of $Q_{ep}$. This interdependence between the two queues results in an interacting system of queues. To decouple this interaction and simplify the analysis, we assume that $Q_p$ is saturated to formulate an expression for the service rate of $Q_{ep}$. The PU is assumed to always have a packet to transmit; this implies that each timeslot an energy packet is consumed from $Q_{ep}$. So, the $Q_{ep}$ service rate, $\mu_{ep} = 1$. So, the probability that $Q_{ep}$ is not empty is $\lambda_{ep}/1$, and the probability that $Q_{ep}$ is empty is $(1 - \lambda_{ep})$. Substituting in (3.2), gives:

$$\mu_p = (\overline{P}_{ps,pd} + P_{ps,pd}\overline{P}_{ps,ss})\lambda_{ep} \tag{3.3}$$

The resulting PU's service rate under the saturation assumption is a lower bound on the actual service rate, therefore the obtained stability region will be an inner bound to the actual stability region.

For the relay queue at the SU, $Q_{ps}$, a packet from the PU enters the relay queue when the channel between the PU transmitter and receiver is in outage, the channel is not in outage between the PU and the SU, the PU battery is not empty, and the PU data queue is not empty, therefore,

$$\lambda_{ps} = P_{ps,pd}\overline{P}_{ps,ss}\, \lambda_{ep}\, \frac{\lambda_p}{\mu_p}. \tag{3.4}$$

The probability that the PU is idle is denoted by $P_I$. the PU is active when both the data queue and the battery queues are non-empty together, otherwise, the PU is idle, hence,

$$P_I = 1 - \lambda_{ep}\,(\lambda_p/\mu_p). \tag{3.5}$$

In our model, randomized cooperative policy, the SU transmits a packet from $Q_s$ or $Q_{ps}$ with probabilities $a$ and $\bar{a}$, respectively. In [37], a comparison between two types of service policy, the full priority service policy to the relay queue and the randomized cooperative policy, is illustrated. It was shown that the randomized cooperative policy enhances the SU delay at the expense of a slight degradation in the PU delay. Therefore, we chose the randomized cooperative policy as the cooperation model between the two energy harvesting primary and secondary users. A packet is serviced from $Q_{ps}$, with a probability $\bar{a}$ if the SU data queue is non-empty, or with a probability one if the SU data queue is empty. As a result, $\mu_{ps}$ can then be expressed as,

$$\mu_{ps} = \overline{P}_{ss,pd}P_{(Q_{es}\neq 0)}P_I\left\{\bar{a}P_{(Q_s\neq 0)} + P_{(Q_s=0)}\right\}, \tag{3.6}$$

where $P_{(Q_s=0)} = 1 - \lambda_s/\mu_s$.

Similarly, a packet is serviced from $Q_s$, with probability $a$ if the SU relay queue is non-empty, or with a probability one if the SU relay queue is empty. Consequently, $\mu_s$ can then be expressed as,

$$\mu_s = \overline{P}_{ss,sd} P_{(Q_{es} \neq 0)} P_I \left\{ a P_{(Q_{ps} \neq 0)} + P_{(Q_{ps} = 0)} \right\}. \tag{3.7}$$

From (3.6) and (3.7), the service rate of the relay queue depends on the current state of the SU data queue and vice versa. Therefore, the two queues are interacting and the individual departure processes cannot be computed directly. Therefore, we resort to the dominant system approach to decouple this interaction [38-39]. In [40], the stability region of the slotted ALOHA system is characterized depending on the stochastic dominance technique. The system was composed of two-node case over a collision channel where nodes are subject to energy availability constraints.

A dominant system has the property that it is stable if and only if the original system is stable, and that its queues are not interacting. The dominant system can be determined by this simple modification to the original system: if $Q_{ps}$ (or $Q_s$) is empty; the SU continues to transmit "dummy" packets whenever it senses the PU is idle. If the SU transmits a dummy packet from $Q_s$ (dominant system *I*), then, $P_{(Q_s \neq 0)} = 1$ and $P_{(Q_s = 0)} = 0$, so from (3.6), a packet is transmitted from $Q_{ps}$ with probability $\bar{a}$ regardless of the actual state of $Q_s$. Conversely, if the SU transmits a dummy packet from $Q_{ps}$ (dominant system *II*), then, $P_{(Q_{ps} \neq 0)} = 1$ and $P_{(Q_{ps} = 0)} = 0$, so from (3.7), a packet is transmitted from $Q_s$ with probability $a$ regardless the actual state of $Q_{ps}$. So, in the two dominant systems, $Q_s$ and $Q_{ps}$ are decoupled and the service rates of $Q_s$ and $Q_{ps}$ could be computed directly. The stable throughput region of the original system would be the union of stable throughput region of the two dominant systems

For a packet to be transmitted from the SU data queues ($Q_{ps}$ or $Q_s$), it is served by an energy packet from $Q_{es}$. The probability of $Q_{es}$ being non-empty is $\lambda_{es}/\mu_{es}$. Since the SU is assumed to transmit dummy packets (from $Q_{ps}$ or $Q_s$), the service rate, $\mu_{es}$, of $Q_{es}$ is the probability that the PU is idle. The probability of $Q_{es}$ being non-empty is $\lambda_{es}/\mu_{es}$, and

$$P_{(Q_{es}\neq 0)} = \lambda_{es}/(1 - \lambda_{ep}(\lambda_p/\mu_p)). \qquad (3.8)$$

Here, the individual departure processes will be computed in the two dominant systems. The stability of the queues under the two dominant systems will be investigated in the next two sections, and the two stability regions of the two systems will be expressed, stability region (*I*) and region (*II*). The stability of the original system is the union of stability region (*I*) and region (*II*).

### 3.2.3 Dominant System *(I)*

The SU is assumed to transmit dummy packets from $Q_s$, so the service rate of the relay queue is independent of the state of $Q_s$. A packet is served from $Q_{ps}$ with probability $\bar{a}$ if the PU is idle, the channel between SU source and the PU destination is not in outage, and $Q_{es}$ not empty, therefore,

$$\mu_{ps} = \overline{P}_{ss,pd} \frac{\lambda_{es}}{1 - \lambda_{ep}(\lambda_p/\mu_p)} P_I \, \bar{a} \qquad (3.9)$$

For the SU relay queue stability, equation (3.1) requires that

$$P_{ps,pd}\overline{P}_{ps,ss}\,\lambda_{ep}\,\frac{\lambda_p}{\mu_p} < \overline{P}_{ss,pd}\frac{\lambda_{es}}{1 - \lambda_{ep}(\lambda_p/\mu_p)}\,P_I\,\bar{a},$$

$$\lambda_p < \frac{\bar{a}\,\overline{P}_{ss,pd}}{P_{ps,pd}\overline{P}_{ps,ss}\,\lambda_{ep}}\,\frac{\lambda_{es}}{1 - \lambda_{ep}(\lambda_p/\mu_p)}\,P_I\,\mu_p. \qquad (3.10)$$

It is clear from the above relations that the maximum allowable arrival rate at $Q_p$ increases as $\bar{a}$ increases, which means that the SU increases the probability of serving the relay queue. Another important note is that this increase implies that the SU has enough energy packets to serve the relay queue. This interprets (3.10) which shows that $\lambda_p$ increases as $\lambda_{es}$ increases. From (3.7) and with the relations: $P_{(Q_{ps}\neq 0)} = \lambda_{ps}/\mu_{ps}$ and $P_{(Q_{ps}=0)} = 1 - \lambda_{ps}/\mu_{ps}$, it can be shown that,

$$\mu_s = \overline{P}_{ss,sd}\frac{\lambda_{es}}{1 - \lambda_{ep}(\lambda_p/\mu_p)}\,P_I\left\{1 - \bar{a}\,\frac{\lambda_{ps}}{\mu_{ps}}\right\} \qquad (3.11)$$

Substituting with $\mu_{ps}$ from (3.9) in (3.11), it is noted that the resulting $\mu_s$ is independent of $\bar{a}$,

$$\mu_s = \overline{P}_{ss,sd} \frac{\lambda_{es}}{1 - \lambda_{ep}(\lambda_p/\mu_p)} P_I \left\{ 1 - \frac{P_{ps,pd}\overline{P}_{ps,ss}\lambda_{ep}\frac{\lambda_p}{\mu_p}}{\overline{P}_{ss,pd}\frac{\lambda_{es}}{1-\lambda_{ep}(\lambda_p/\mu_p)}P_I} \right\}. \quad (3.12)$$

This can be explained as follows. when $a$ increases, the first term in the brackets of (3.7), $aP_{(Q_{ps}\neq 0)}$, would increase, and the probability of $Q_{ps}$ being empty, would decrease, which is the second term in the brackets of equation (3.7). So, changing the access probability $a$ has no effect on the SU service rate, $\mu_s$, in dominant system (*I*). From (3.1), (3.10), (3.12), the stability region of system (*I*) would be bounded by:

$$R_I = \begin{cases} (\lambda_p, \lambda_s): \lambda_p < \mu_p, \\ \lambda_p < \frac{\bar{a}\,\overline{P}_{ss,pd}}{P_{ps,pd}\overline{P}_{ps,ss}\lambda_{ep}} \frac{\lambda_{es}}{1-\lambda_{ep}(\lambda_p/\mu_p)} P_I \mu_p, \\ \lambda_s < \overline{P}_{ss,sd} \frac{\lambda_{es}}{1-\lambda_{ep}(\lambda_p/\mu_p)} P_I \left\{ 1 - \frac{P_{ps,pd}\overline{P}_{ps,ss}\lambda_{ep}\frac{\lambda_p}{\mu_p}}{\overline{P}_{ss,pd}\frac{\lambda_{es}}{1-\lambda_{ep}(\lambda_p/\mu_p)}P_I} \right\} \end{cases}. \quad (3.13)$$

In Fig. 3.2, the achievable stable throughput region of dominant system (*I*) is plotted. Hereafter, the system parameters are chosen as follows: $\overline{P}_{ps,pd} = 0.3$, $\overline{P}_{ps,ss} = 0.4$, $\overline{P}_{ss,pd} = 0.7$, and $\overline{P}_{ss,sd} = 0.7$. The arrival rates of the two battery queues are as follow: $\lambda_{ep} = 0.6$, $\lambda_{es} = 0.6$. By examining the horizontal axis ($\lambda_p$) it is clear that the maximum sustainable $\lambda_p$ decreases as $a$ increases, when the SU tends to serve its data queue at the expense of the relay queue. First, as $\lambda_p$ increases, probability of PU being idle, $P_I$, decreases and $P_{(Q_{es}\neq 0)}$ increases with the same amount as $P_I$. Concurrently, the maximum sustainable $\lambda_s$ decreases as $\lambda_{ps}$ increases from (3.11). Second, as $\lambda_p$ increases, $P_{(Q_{es}\neq 0)}$ reaches its maximum of unity, $P_I$ decreases and $\lambda_{ps}$ increases. From (3.11), the rate of decrease of $\lambda_s$ would be faster as $\lambda_{ps}$ increases and $P_I$ decreases. This occurs when $\mu_{es} = 1 - \lambda_{ep}(\lambda_p/\mu_p) = \lambda_{es}$ and $\lambda_p = 0.25$. Third, the maximum sustainable $\lambda_p$ from (3.3) is $\mu_p = 0.34$.

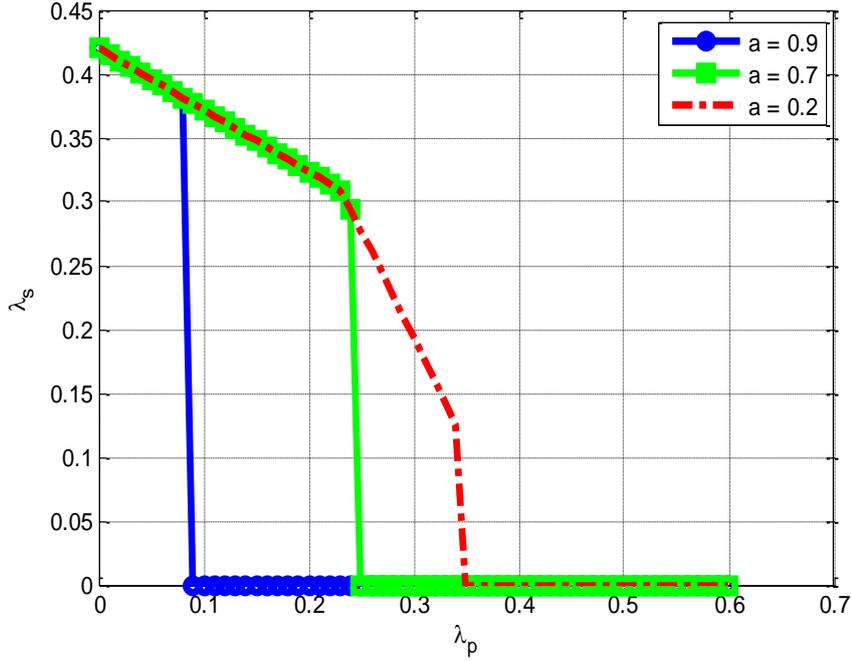

**Figure 3.2**  Stable throughput region for dominant system (*I*) at different values of service probability $a$ ($\lambda_{ep} = 0.6, \lambda_{es} = 0.6$)

### 3.2.3 Dominant System *(II)*

In this dominant system, dummy packets transmission from $Q_{ps}$ makes the service rate of the SU data queue decoupled from the current state of $Q_{ps}$. A packet is served from $Q_s$ with probability $a$ if the PU is idle, channel between SU source and destination not in outage, and $Q_{es}$ not empty. For the SU data queue service rate,

$$\mu_s = \overline{P}_{ss,sd} \frac{\lambda_{es}}{1 - \lambda_{ep}(\lambda_p/\mu_p)} P_I a \tag{3.14}$$

From (3.6) and with the relations $P_{(Q_s \neq 0)} = \lambda_s/\mu_s$ and $P_{(Q_s = 0)} = 1 - \lambda_s/\mu_s$, it can be verified that:

$$\mu_{ps} = \overline{P}_{ss,pd} \frac{\lambda_{es}}{1 - \lambda_{ep}(\lambda_p/\mu_p)} P_I \left\{1 - a\frac{\lambda_s}{\mu_s}\right\} \tag{3.15}$$

Substituting (3.14) in (3.15),

$$\mu_{ps} = \overline{P}_{ss,pd} \frac{\lambda_{es}}{1 - \lambda_{ep}(\lambda_p/\mu_p)} P_I \left\{ 1 - \frac{\lambda_s}{\overline{P}_{ss,sd} \frac{\lambda_{es}}{1 - \lambda_{ep}(\lambda_p/\mu_p)} P_I} \right\}. \qquad (3.16)$$

It is noted that as $\bar{a}$ increases, the first term in the brackets of (3.6), $\bar{a}P_{(Q_s \neq 0)}$, would increase, and the probability of $Q_s$ being empty would decrease, which is the second term in the brackets of (3.6). Therefore, changing the access probability $\bar{a}$, has no effect on the service rate of $Q_{ps}$ in the dominant system (*II*). For the SU relay queue stability, with $\mu_{ps}$ derived in (3.16), it can be verified that,

$$\lambda_{ps} < \mu_{ps},$$

$$\lambda_p < \frac{\mu_p/\lambda_{ep}}{\overline{P}_{ss,pd} P_{(Q_{es} \neq 0)} + P_{ps,pd} \overline{P}_{ps,ss}} \times \left( \overline{P}_{ss,pd} P_{(Q_{es} \neq 0)} - \frac{\lambda_s}{C} \right), \qquad (3.17)$$

where $C = \frac{\overline{P}_{ss,sd}}{\overline{P}_{ss,pd}}$. From (3.1), (3.14), (3.17), the stability region of dominant system (*II*) would be bounded by

$$R_{II} = \left\{ \begin{array}{c} (\lambda_p, \lambda_s): \lambda_p < \mu_p, \\ \lambda_s < \overline{P}_{ss,sd} \frac{\lambda_{es}}{1 - \lambda_{ep}(\lambda_p/\mu_p)} P_I a, \\ \lambda_p < \frac{\mu_p/\lambda_{ep}}{\overline{P}_{ss,pd} P_{(Q_{es} \neq 0)} + P_{ps,pd} \overline{P}_{ps,ss}} \\ \times \left( \overline{P}_{ss,pd} P_{(Q_{es} \neq 0)} - \frac{\lambda_s}{C} \right) \end{array} \right\}. \qquad (3.18)$$

Figure 3.3 depicts the stability region under dominant system (*II*) for different values of service probability $a$. As shown in (3.16), the service rate of $Q_{ps}$ is independent of $a$. The PU maximum allowable arrival rate ($\lambda_p$) depends only on $Q_p$ service rate, from (3), and $Q_{ps}$ service rate, from (16). So, the PU maximum allowable arrival rate ($\lambda_p$) rate is independent of $a$. By observing the vertical axis ($\lambda_s$), the maximum sustainable ($\lambda_s$) increases as $a$ increases, where the SU tends to serve $Q_s$ with higher service probability.

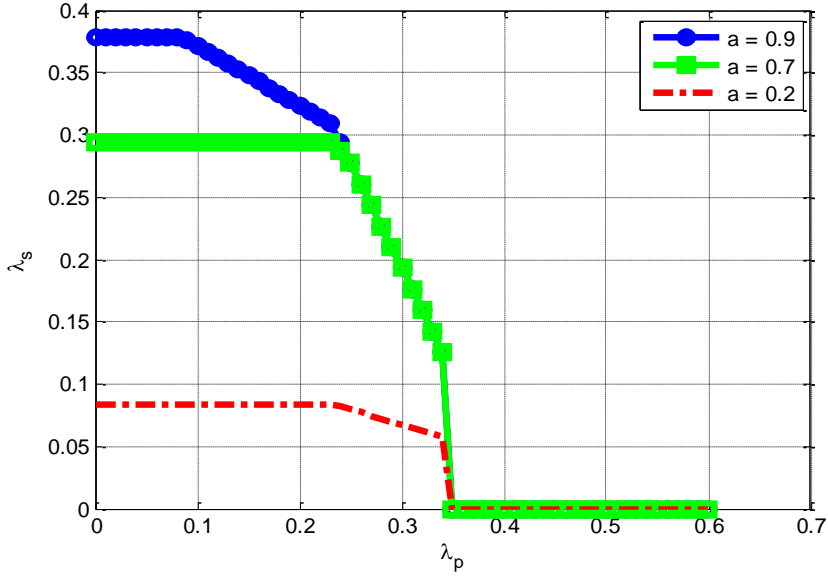

**Figure 3.3**  Stable throughput region for dominant system (*II*) at different values of service probability $a$ ($\lambda_{ep} = 0.6, \lambda_{es} = 0.6$)

## 3.2.4 Performance Analysis

In this section, the performance of the cooperative cognitive system under the energy constraint is investigated. The stability region of the discussed system is defined as the union of the two stable throughput regions of dominant system (*I*) and dominant system (*II*), $R = R_1 \cup R_2$. Hereafter, the system parameters are chosen as follows: $\overline{P}_{ps,pd} = 0.3, \overline{P}_{ps,ss} = 0.4, \overline{P}_{ss,pd} = 0.7$, and $\overline{P}_{ss,sd} = 0.7$.

In Fig. 3.4, the stability region of the overall system R is plotted. It is worth mentioning that, the maximum allowable arrival rates for the PU and SU are independent of $a$ in R since it is the union of the two dominant systems, $R_1 \cup R_2$, over all values of $a$.

In Fig. 3.5, the system without energy constraint is compared to energy constrained system with $\lambda_{es} = 0.5$ and $\lambda_{ep} = 1$, i.e. energy constraint on the SU only. With the arrival rate at the SU battery queue is halved, the maximum sustainable SU throughput is $(\lambda_{es}\overline{P}_{ss,sd}) = 0.35$.

It is intuitive that the energy constrained system ($\lambda_{es} = 0.5$) is always lower than the original system ($\lambda_{es} = 1$). This is true up to certain $\lambda_p$, after which the boundaries of the two stability regions coincide. This can be understood as from (3.8), the probability of $Q_{es}$ being non-empty is ($\lambda_{es}/P_I$). As $\lambda_p$ increases, the probability that PU is idle decreases and the probability of $Q_{es}$ being non-empty gradually approaches one. This makes the energy limited system act as the original system since in both systems $P_{(Q_{es} \neq 0)} = 1$. This value of $\lambda_p$ is reached when $\lambda_{es} = P_I = 1 - (1)(\lambda_p/\mu_p) = 0.5$ and $\mu_p = 0.58$ from (3.3), so, $\lambda_p = 0.29$.

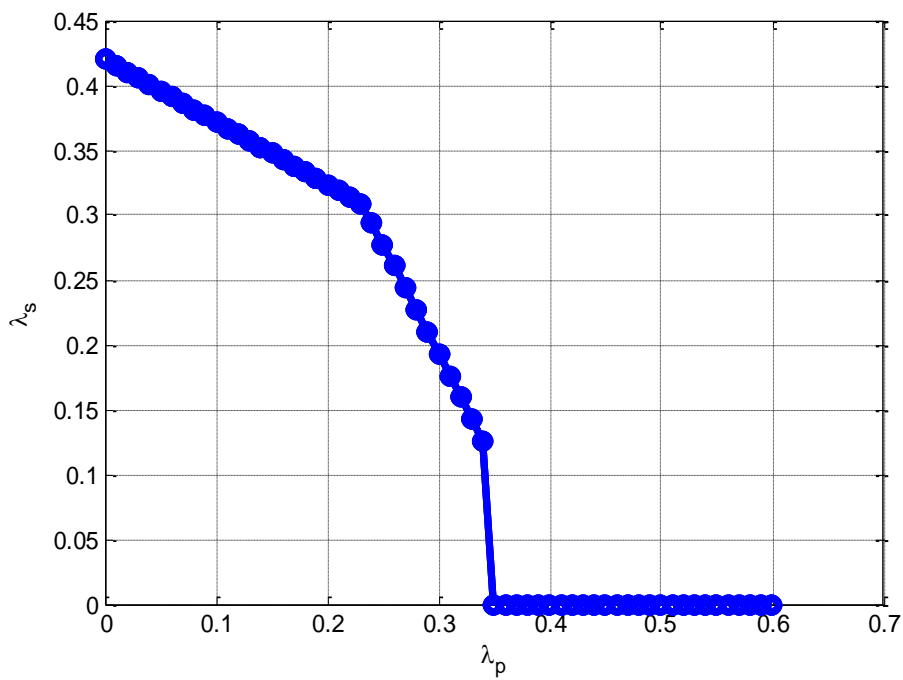

**Figure 3.4  Stable throughput region of the overall system ($\lambda_{ep} = 0.6, \lambda_{es} = 0.6$)**

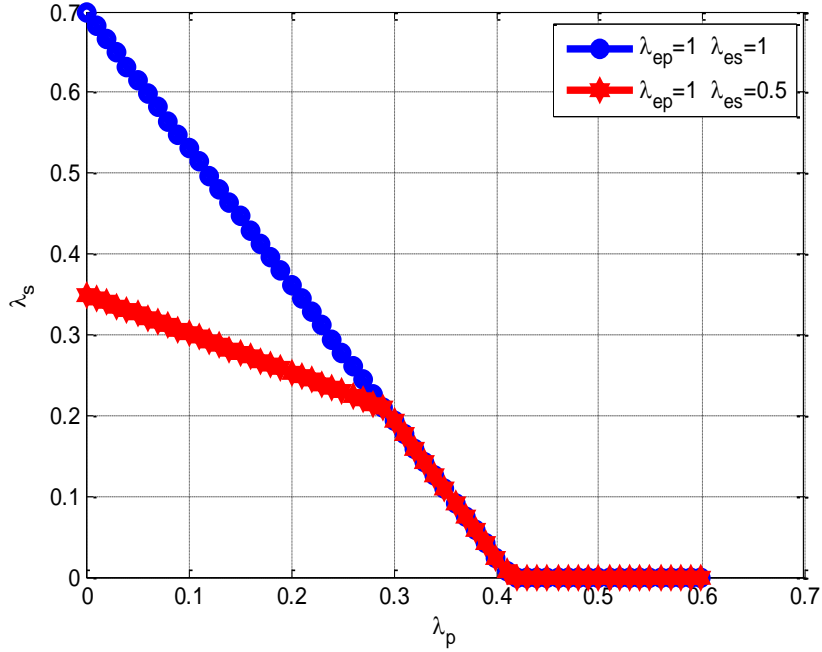

**Figure 3.5    Stable throughput region of the overall system with SU energy harvesting**

In contrast to Fig. 3.5, system with energy constraint on the PU only is shown in Fig. 3.6. The energy constraint at PU is as, $\lambda_{ep} = 0.6$ and no constraint on SU since $\lambda_{es} = 1$. The influence of $\lambda_{ep}$ is directly obtained from (3.3), where $\mu_p$ decreases as $\lambda_{ep}$ decreases. For the stability of $Q_p$, the maximum allowable PU arrival rate is also decreased to maintain the stability condition presented in (3.1). This cut in the horizontal axis ($\lambda_p$) occurs when the PU arrival rate equals the PU service rate, $\lambda_p = \mu_p = 0.6 \times 0.58 = 0.34$.

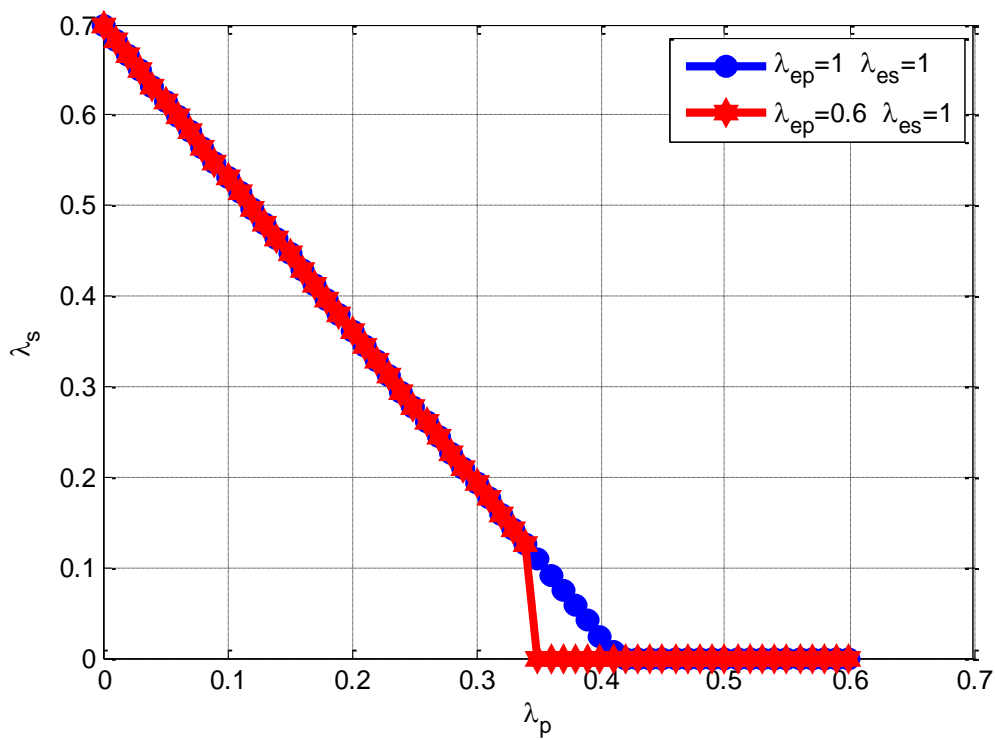

**Figure 3.6    Stable throughput region of the overall system with PU energy harvesting**

Combing the effect of energy limitation on PU and SU, shrinks the stability region in both directions (for PU and SU) as in Fig. 3.7, it is seen that the two boundaries coincide when the PU activity increases and SU battery queue is able to cover transmission on the available timeslots. In Fig. 3.8, this mentioned coincidence does not exist since with more severe energy limitation, the PU probability of being idle is higher and no sufficient energy at the SU to exploit these idle timeslots.

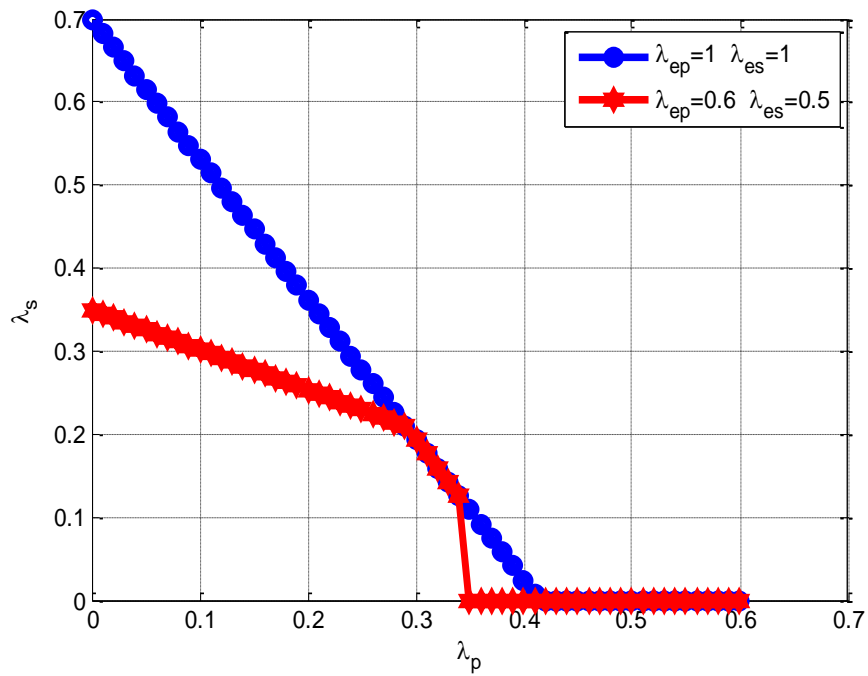

**Figure 3.7    Stable throughput region of the overall system with PU and SU energy harvesting (moderate harvesting rates)**

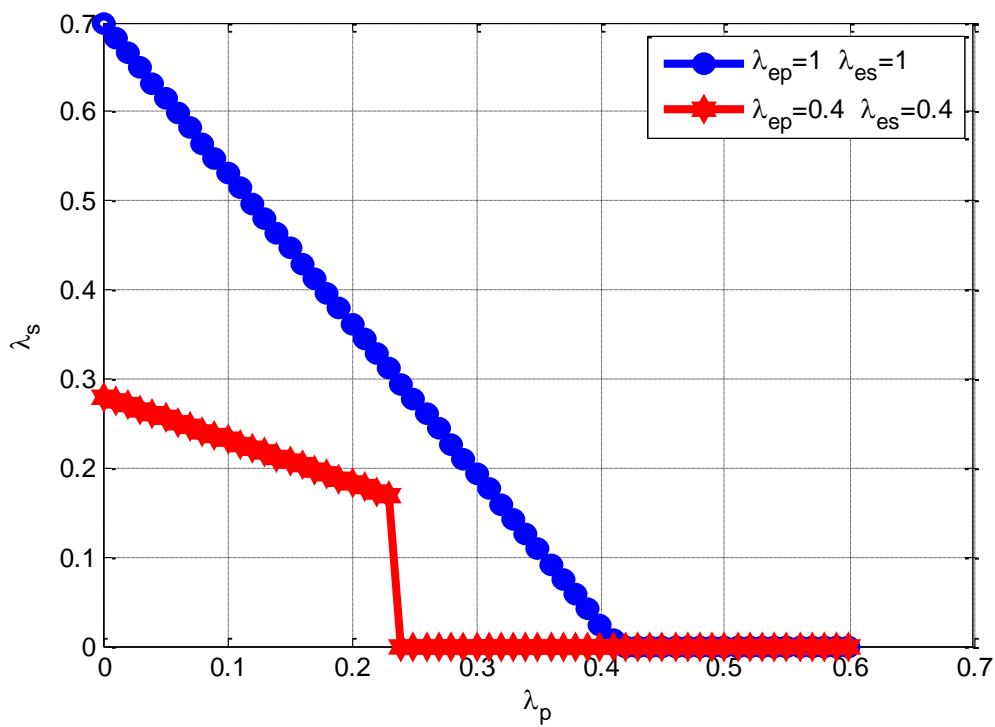

**Figure 3.8    Stable throughput region of the overall system with PU and SU energy harvesting (low harvesting rates)**

## 3.2.6 Non-cooperative energy harvesting system

For the sake of comparison, we introduce the energy harvesting cognitive radio system without cooperation between PU and SU. For $Q_p$, a packet is serviced if it is successfully decoded by the PU destination only. With the same derivations of service rates of battery queues and $P_I$, we can get:

$$\mu_p = \overline{P}_{ps,pd}\, \lambda_{ep}, \qquad (3.19)$$

$$\mu_s = \overline{P}_{ss,sd}\, \frac{\lambda_{es}}{1 - \lambda_{ep}(\lambda_p / \mu_p)}\, P_I. \qquad (3.20)$$

The stability regions of the two systems (with and without cooperation) are plotted in Fig. 3.9 [19] [24]. The same system parameters as before, also, the arrival rates of the two battery queues are $\lambda_{ep} = 1$, $\lambda_{es} = 1$. It is clear that for the two systems, the maximum sustainable SU throughput is $(\lambda_{es}\overline{P}_{ss,sd})$. Cooperative system provides better SU throughput for all possible values of $\lambda_p$ also, the maximum sustainable PU arrival rate, $\lambda_p$, is larger in the cooperative system. This means that cooperation between PU and SU is beneficial for both users. Taking into account the effect of energy harvesting, $\lambda_{ep}$ and $\lambda_{es}$ should be considered less than unity.

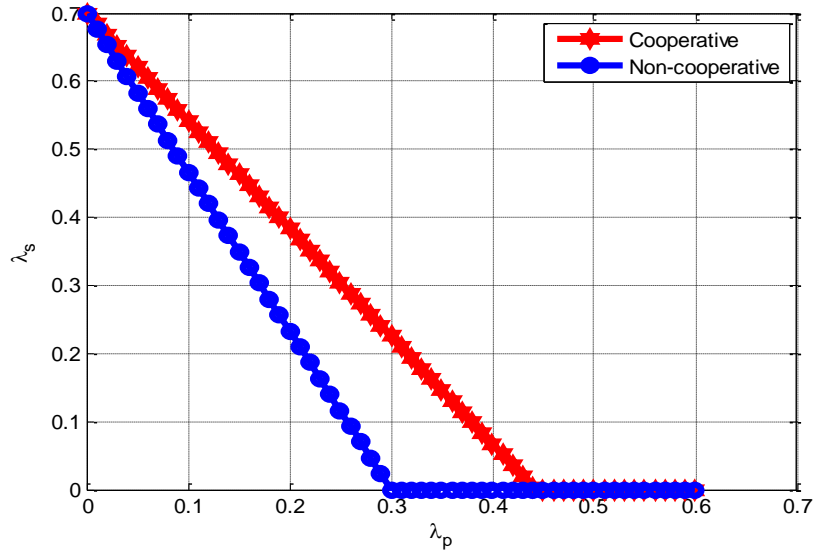

**Figure 3.9    Stable throughput region of the overall cooperative system and non-cooperative system**

The stability regions of the two systems (with and without cooperation) are plotted in Fig. 3.10 and Fig. 3.11 with $\lambda_{ep} = 0.5$, $\lambda_{es} = 0.8$ and $\lambda_{ep} = 0.5$, $\lambda_{es} = 0.6$, respectively. Fig. 3.10 and Fig. 3.11 show that non-cooperative energy harvesting system is better than cooperative energy harvesting system for low values of $\lambda_p$ in terms of SU throughput. At certain $\lambda_p$, the two systems intersect providing the same SU allowable throughput, after which the cooperative system is better in terms of SU throughput and maximum allowable PU arrival rate.

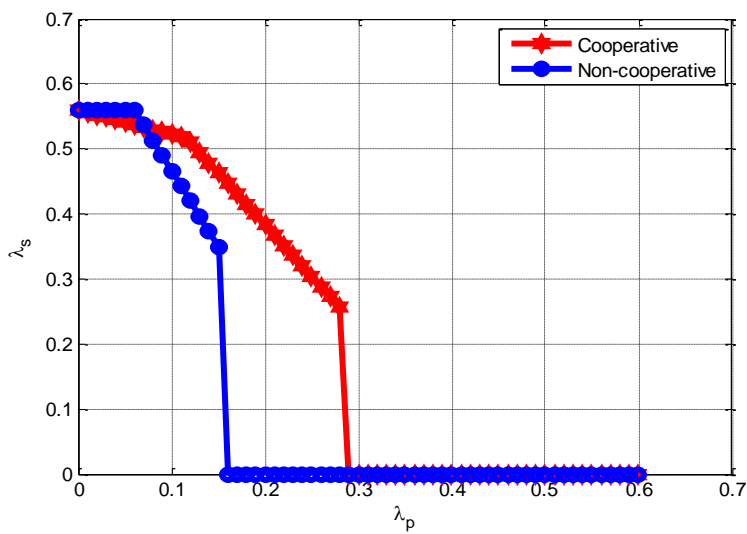

**Figure 3.10    Stable throughput region of the overall system with PU and SU energy harvesting ($\lambda_{ep} = 0.5, \lambda_{es} = 0.8$)**

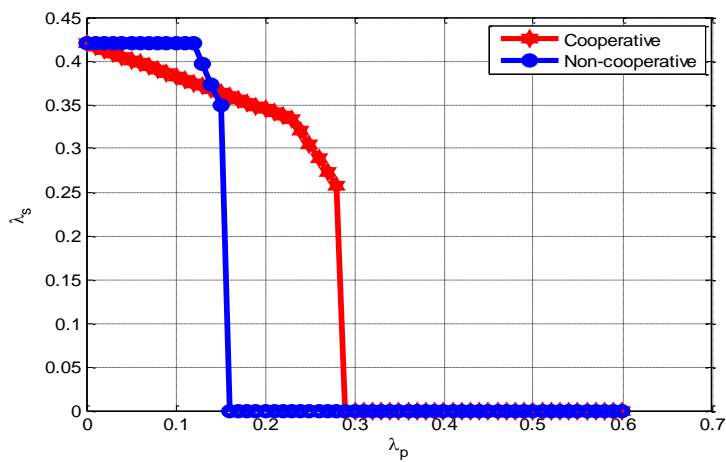

**Figure 3.11    Stable throughput region of the overall system with PU and SU energy harvesting ($\lambda_{ep} = 0.5, \lambda_{es} = 0.6$)**

To obtain the value of $\lambda_p$ after which the energy harvesting cooperative system became superior to the non-cooperative system, we would equate $\mu_s$ from (3.12) with $\mu_s$ from (3.20). After some algebraic manipulation, we get

$$\overline{P}_{ss,sd} \frac{\lambda_{es}}{1 - \lambda_{ep}(\lambda_p/\mu_p)} P_I \left\{ 1 - \frac{P_{ps,pd}\overline{P}_{ps,ss}\lambda_{ep}\frac{\lambda_p}{\mu_p}}{\overline{P}_{ss,pd}\frac{\lambda_{es}}{1-\lambda_{ep}(\lambda_p/\mu_p)}P_I} \right\}$$

$$= \overline{P}_{ss,sd} \frac{\lambda_{es}}{1-\lambda_{ep}(\lambda_p/\mu_p)} P_I,$$

$$\Lambda_p = \frac{1-\lambda_{es}}{D}, \qquad (3.21)$$

where $\Lambda_p$ is the PU arrival rate at which the two systems (cooperative and non-cooperative) are with the same SU allowable throughput, and D is a constant function of the channel outages probabilities

$$D = \frac{1}{\overline{P}_{ps,pd}} - \frac{P_{ps,pd}\overline{P}_{ps,ss}}{\overline{P}_{ss,pd}(\overline{P}_{ps,pd} + P_{ps,pd}\overline{P}_{ps,ss})}. \qquad (3.22)$$

From (22), it is clear that the value, $\Lambda_p$, at which the two systems intersect is a function of the system channel parameters and $\lambda_{es}$. With the mentioned values of $\lambda_{es}$, $\Lambda_p = 0.075$ for Fig. 3.10 and $\Lambda_p = 0.15$ for Fig. 3.11. It is clear that this value of $\Lambda_p$ decreases as $\lambda_{es}$ increases. This result can be interpreted as follows: the SU with higher energy levels at its battery is supposed to has better opportunity to cooperate with the PU and leverage the better channel conditions between the SU and PU destination. Figure 3.12 shows the three dimensional plot of the intersection point $\Lambda_p$ versus $\lambda_{ep}$ and $\lambda_{es}$. We can interpret Fig. 3.12 as follows: when $\lambda_{es} = 0$, $\Lambda_p$ is always zero indicating that the cooperative method has no benefits due to lack of energy at the SU. When $\lambda_{es} = 1$, $\Lambda_p$ is always zero indicating that the cooperative method is outperforming the non-cooperative method. Finally, as $\lambda_{ep}$ increases, the required value of $\lambda_{es}$ to make the cooperation outperforming the non-cooperation also increases.

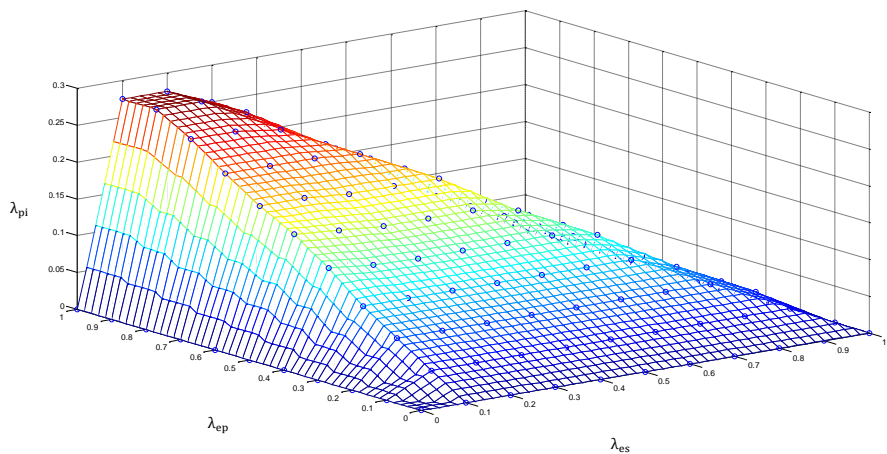

**Figure 3.12　Three dimensional plot of $\Lambda_p$ versus $\lambda_{ep}$ and $\lambda_{es}$**

## 3.3 MULTI SECONDARY USER ENERGY HARVESTING NETWORK MODEL

As an extension to the system model introduced in Fig. 3.1 and the associated analysis of in the previous sections, we introduce here the system of multi-SU nodes with energy harvesting cooperates with one PU with energy harvesting. We go through almost the same assumptions and theorems used in the previous analysis.

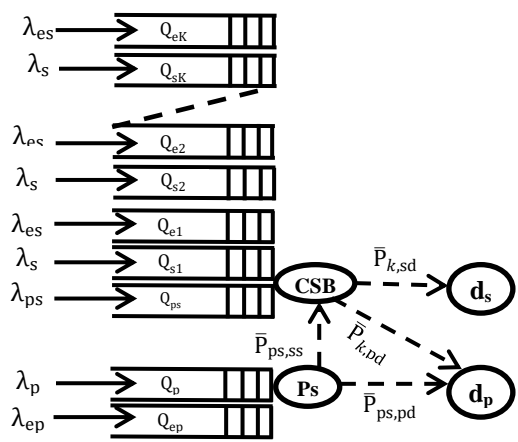

**Figure 3.13　System model of Multi-SU cluster**

## 3.3.1 System Configuration

Figure 3.13 depicts system model under consideration. The system consists of one primary user (PU) and a cluster of SUs, $S_{\text{relay}} = \{1, 2, ...., K\}$ with $K$ nodes. For convenience, we assume that $K$ secondary nodes are clustered relatively close together (location-based clustering). The system of one PU and $K$-SUs is assumed to harvest energy from the environment. $Q_k$ is the queue of SU node $k \in S_{\text{relay}}$. The cognitive cluster is equipped with a common relaying queue $Q_{\text{ps}}$ which is used for cooperation and is accessible from all secondary nodes. A cluster supervision block is used to synchronize and control all the activities of the cognitive cluster. This central logic retains the common queue and is the interface between the cognitive cluster and the primary network.

It is assumed that secondary nodes can perfectly exchange information with the CSB without overhead. Regarding the SU cluster, each SU transmitter $k \in S_{\text{relay}}$ is equipped with $Q_{ek}$ queue, where $Q_{ek}$ is the SU transmitter battery storing the harvested energy. The arrival processes at the two queue groups, $Q_k$ and $Q_{ek}$, are modeled as Bernoulli arrival process with means $\lambda_s$ and $\lambda_{es}$, respectively for each cognitive user.

The PU transmits a packet from $Q_p$ whenever it is non-empty. If the channel between the PU transmitter and receiver is not in outage, then the PU receiver successfully decodes the packet and the packet departs the system. It is assumed that the SU cluster can overhear the ACK/NACK from the PU receiver. In timeslots when the channel between the PU transmitter and receiver is in outage, the PU's packet is added to the common relaying queue if the PU it is decoded by at least one secondary node of the cognitive cluster. The cluster broadcasts an ACK signal to the PU indicating that SU cluster will bear the responsibility to deliver this packet.

If the channel between the PU and SU also is in outage, PU will try to retransmit the packet in a subsequent timeslot. All cognitive nodes, which successfully decode the source message, convey their packets to the central CSB. After a basic processing of the received packets and discarding replicated packets, the CSB enters the source packet to the common queue and transmits an ACK signal. According to the previous description of the cognitive structure, the relay cluster can decode the transmitted signal if at least one relay can decode it. Whenever the channel is sensed to be idle, the secondary cluster has two options for packet transmission, specifically from any secondary node $Q_k$ or from $Q_{ps}$. The CSB controls the selection of a packet for transmission, whether to transmit a packet from $Q_k$ with probability $a$, or from $Q_{ps}$ with the complement probability $\bar{a}$. The selection of the secondary node is based on the quality of the link between the selected node and the secondary destination.

### 3.3.2 Stability Regions

This section describes the stability region for the system presented in the previous subsection and depicted in Fig. 3.13. This region is bounded by the maximum arrival rates at the PU and each SU node when all the queues of the system are stable. For any queue in the system, the stability requires that: $\lambda_i < \mu_i$, where $i \in \{p, ps, s, ep, es\}$, and $\mu_i$ refers to the service rate of the $i^{th}$ queue. A packet from $Q_p$ is served if $Q_{ep}$ is not empty and it is successfully decoded by the PU destination or by at least one secondary node of the cognitive cluster. We then have,

$$\mu_p = (\overline{P}_{ps,pd} + P_{ps,pd}\overline{P}_{ps,k*}) P_{(Q_{ep} \neq 0)}. \tag{3.23}$$

where $k^*$ is chosen such that, the channel gain between the PU source and the SU-cluster is maximized, so $k^* = \text{argmax}_{k \in S_{relay}}\{\gamma_{ps,k}\}$, where $\gamma_{ps,k}$ is the channel gain between the PU source and the $k^{th}$ SU of the SU cluster. It is given by order statistics [41] that, $\overline{P}_{ps,k*} = 1 - [1 - \overline{P}_{ps,k}]^K = 1 - [P_{ps,k}]^K$. Probability of having $Q_{ep}$ empty is driven from Little's law by $(1 - \lambda_{ep}/\mu_{ep})$. Based on that, the service rate of $Q_p$ depends on whether $Q_{ep}$ is empty or not and vice versa as in the single SU system model.

We resort to the same assumption as before to simplify the analysis of this interacting system of queues by assuming that $Q_p$ is saturated to decouple the interaction with $Q_{ep}$ and formulate an expression for $\mu_p$. So, the PU is assumed to have a packet to transmit in each timeslot; this implies that each timeslot an energy packet is dissipated from $Q_{ep}$, and $\mu_{ep} = 1$. The PU's service rate derived under the saturation assumption is a lower bound on the actual service rate of the system. Therefore, the resulting stability region is an inner bound to the actual stability region of the system.

Considering the SU CSB, a packet is admitted to the relay queue when the channel between the PU transmitter and receiver is in outage, concurrently the PU packet is decoded by at least one secondary node of the cognitive cluster, the PU data queue is not empty, and the PU battery is not empty, then,

$$\lambda_{ps} = P_{ps,pd}\overline{P}_{ps,k*}\,\lambda_{ep}\,\frac{\lambda_p}{\mu_p}. \tag{3.24}$$

The SU cluster CSB decides to transmit a packet from $Q_k$, $k \in S_{\text{relay}}$, or $Q_{ps}$ whenever the PU is idle and $Q_{ek}$ is non-empty. A packet is serviced from $Q_{ps}$, with a probability $\bar{a}$ if at least one SU node $k \in S_{\text{relay}}$ of the SU cluster is non-empty, or with a probability one if all SU cluster is empty. The SU cluster is defined to be empty if all data queues, $Q_k$, $k \in S_{\text{relay}}$, are empty which occurs with probability $P_{(Q_s=0)}$ and conversely, it is defined as non-empty if at least one queue of the cluster is non-empty which occurs with probability $P_{(Q_s \neq 0)}$. $\mu_{ps}$ can then be expressed as,

$$\mu_{ps} = \overline{P}_{k1,pd}P_{(Q_{es}\neq 0)}\left\{\bar{a}P_{(Q_s\neq 0)} + P_{(Q_s=0)}\right\}P_I, \tag{3.25}$$

where $k1 = \text{argmax}_{k\in S_{\text{relay}}}\{\gamma_{k,pd}\}$, For a given $k$-SU of the SU cluster, $\overline{P}_{k1,pd}=1-[1-\overline{P}_{k,pd}]^K = 1-[P_{k,pd}]^K$. Similarly, a packet is serviced from certain $Q_k$ with a probability $a$ if this queue has the best link to the SU common destination and $Q_{ps}$ is non-empty, or with a probability one if $Q_{ps}$ is empty, $\mu_s$ can then be expressed as,

$$\mu_s = \Delta_k\,\overline{P}_{k2,sd}P_{(Q_{es}\neq 0)}\left\{aP_{(Q_{ps}\neq 0)} + P_{(Q_{ps}=0)}\right\}P_I, \tag{3.26}$$

where $\Delta_k$ denotes the event that relay $k$ is selected for transmission ($k = k2$) and $k2 = \text{argmax}_{k \in S_{\text{relay}}}\{\gamma_{k,\text{sd}}\}$. For a given $k$-SU of the SU cluster, $\overline{P}_{k2,\text{sd}} = 1 - [1 - \overline{P}_{k,\text{sd}}]^K = 1 - [P_{k,\text{sd}}]^K$.

We note that due to the clustered configuration of the secondary nodes, the average channels between them and the secondary destination are assumed to be statistically equivalent [42]. Therefore, each relay can access the channel with the same probability, $\Delta_k = 1/K$, and this behavior yields long-term transmission fairness among the SU nodes.

Due to the interaction between $Q_{\text{ps}}$ and $Q_s$ from (3.25) and (3.26), $\mu_s$ and $\mu_{\text{ps}}$ cannot be computed directly. We bypass this difficulty by utilizing the idea of stochastic dominance approach. The stability region of the dominant system is an inner-bounds that of the original system. The dominant system can be obtained from the original system by this modification: if $Q_{\text{ps}}$ (or all $Q_k$, $k \in S_{\text{relay}}$) is empty, the SU CSB decides to continue transmission of "dummy" packets whenever it senses the PU is idle. ($Q_s$ is empty means that all $Q_k$ are empty and conversely, $Q_s$ is non-empty means that at least there is one $Q_k$ being non-empty.)

If the SU cluster transmits a dummy packet from $S_{\text{relay}}$ (dominant system *I*), then, $P_{(Q_s \neq 0)} = 1$ and $P_{(Q_s = 0)} = 0$. Therefore, from (3.6), a packet is transmitted from $Q_{\text{ps}}$ with probability $\bar{a}$ regardless of the actual state of $S_{\text{relay}}$. Conversely, if the SU cluster transmits a dummy packet from $Q_{\text{ps}}$ (dominant system *II*), then, $P_{(Q_{\text{ps}} \neq 0)} = 1$ and $P_{(Q_{\text{ps}} = 0)} = 0$. Therefore, from (3.7), a packet is transmitted from $S_{\text{relay}}$ with probability $a$ regardless the actual state of $Q_{\text{ps}}$.

If a packet is transmitted from the SU cluster nodes or $Q_{\text{ps}}$, it is served by an energy packet from $Q_{ek}$, $k = \{1, \ldots, K\}$. The probability of having any $Q_{ek}$ non-empty is $\lambda_{ek}/\mu_{ek}$. Since the SU network is assumed to transmit dummy packets, the service rate of certain $Q_{ek}$, $\mu_{ek}$, is the probability that the corresponding data queue, $Q_k$, is chosen for transmission when the PU is idle,

$$P_{(Q_{ek} \neq 0)} = \frac{\lambda_{es}}{\Delta_k (1 - \lambda_{ep}(\lambda_p/\mu_p))}. \tag{3.27}$$

In the next two subsections, the individual departure processes will be computed in the two dominant systems and the stability of the queues will be investigated.

### 3.3.3 Dominant System (*I*)

We consider the first dominant system in which the SU cluster transmits dummy packets from any queue $Q_k$ whenever its queues are empty while serving packets from $Q_{ps}$ in the same way as in the original system. All other assumptions remain unaltered in the dominant system. Thus, the service rate at the relay queue is given by:

$$\mu_{ps} = \overline{P}_{k1,pd} \frac{K \lambda_{es}}{1 - \lambda_{ep} (\lambda_p / \mu_p)} P_I \, \bar{a} \tag{3.28}$$

To derive the stability condition for the relay queue, equation (3.1) requires that:

$$P_{ps,pd} \overline{P}_{ps,k*} \lambda_{ep} \frac{\lambda_p}{\mu_p} < \overline{P}_{k1,pd} \frac{K \lambda_{es}}{1 - \lambda_{ep} (\lambda_p / \mu_p)} P_I \, \bar{a}$$

$$\lambda_p < \frac{\bar{a} \, \overline{P}_{k1,pd}}{P_{ps,pd} \overline{P}_{ps,k*} \lambda_{ep}} \frac{K \lambda_{es}}{1 - \lambda_{ep} (\lambda_p / \mu_p)} P_I \, \mu_p . \tag{3.29}$$

It is clear from (3.29) that the maximum sustainable PU arrival rate increases as $\bar{a}$ increases when the SU cluster tends serve the relay queue. Also the maximum PU arrival rate increases with increasing the harvesting rate at the SU nodes and the number of these nodes, ($K \lambda_{es}$). From (3.26), it can be shown that:

$$\mu_s = \overline{P}_{k2,sd} \, \Delta_k \, \frac{K \lambda_{es}}{1 - \lambda_{ep} (\lambda_p / \mu_p)} \left\{ 1 - \bar{a} \frac{\lambda_{ps}}{\mu_{ps}} \right\} P_I \tag{3.30}$$

With $\mu_{ps}$ from (3.28) substituted in (3.30), it is noted that the resulting $\mu_s$ is independent of $a$,

$$\mu_s = \overline{P}_{k2,sd} \, \Delta_k \, \frac{K \lambda_{es}}{1 - \lambda_{ep} (\lambda_p / \mu_p)} P_I \left\{ 1 - \frac{P_{ps,pd} \overline{P}_{ps,k*} \lambda_{ep} \frac{\lambda_p}{\mu_p}}{\overline{P}_{k1,pd} \frac{K \lambda_{es}}{1 - \lambda_{ep} (\lambda_p / \mu_p)} P_I} \right\}. \tag{3.31}$$

When the SU cluster tends to serve its own node queues with a higher service policy, $a$ increases, $P_{(Q_{ps}=0)}$ would decrease and the term $a \, P_{(Q_{ps} \neq 0)}$ would increase.

Consequently, the summation in (3.26), $a\, P_{(Q_{ps}\neq 0)} + P_{(Q_{ps}=0)}$, is constant independent of the service policy $a$.

From (3.1), (3.29), (3.31), the stability region of system (*I*) would be bounded by:

$$R_I = \begin{cases} (\lambda_p, \lambda_s): \\ \lambda_p < \dfrac{\bar{a}\,\overline{P}_{k1,pd}}{P_{ps,pd}\overline{P}_{ps,k*}\lambda_{ep}} \dfrac{K\lambda_{es}}{1-\lambda_{ep}(\lambda_p/\mu_p)} P_I\, \mu_p, \\ \lambda_s < \overline{P}_{k2,sd}\varDelta_k \dfrac{K\lambda_{es}}{1-\lambda_{ep}(\lambda_p/\mu_p)} P_I\left\{1 - \dfrac{P_{ps,pd}\overline{P}_{ps,k*}\lambda_{ep}\frac{\lambda_p}{\mu_p}}{\overline{P}_{k1,pd}\frac{K\lambda_{es}}{1-\lambda_{ep}(\lambda_p/\mu_p)}P_I}\right\} \end{cases}. \quad (3.32)$$

In Fig. 3.14, we show the stability region at three values of service policy $a$ of the dominant system (*I*). Hereafter, the system parameters are chosen as follows: $K = 2$, $\overline{P}_{ps,pd} = 0.3$, $\overline{P}_{ps,k} = 0.4$, $\overline{P}_{k,pd} = 0.8$, and $\overline{P}_{k,sd} = 0.8$. The arrival rates of the battery queues are as follow: $\lambda_{ep} = 0.5$, $\lambda_{es} = 0.5$. It is intuitive that the maximum sustainable $\lambda_p$ decreases as $a$ increases, where the SU CSB tends to serve its own nodes at the expense of the relay queue. Conversely, the maximum SU service rate is independent of the probability $a$, as shown in (3.31). For $K = 2$, the maximum achievable $\lambda_s$ is $\varDelta_k\,(1 - [P_{k,sd}]^K) = 0.47$.

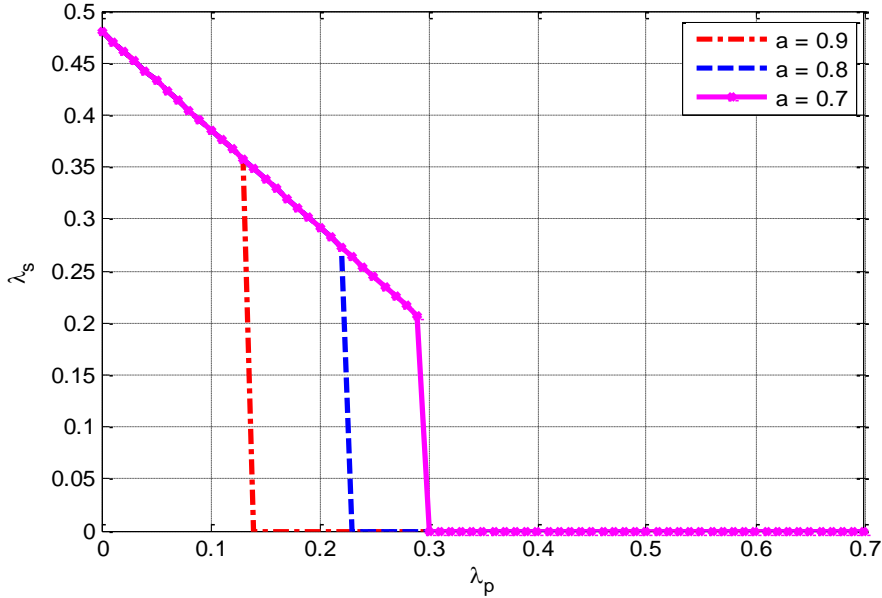

**Figure 3.14    Stable throughput region for dominant system (*I*) at different values of service probability $a$ ($\lambda_{ep} = 0.5, \lambda_{es} = 0.5$)**

### 3.3.4 Dominant System (*II*)

Here we consider the second dominant system, in which the SU transmits dummy packets from the relay queue whenever the relay queue is empty while serving packets from SU nodes in is the same way as in the original system. All other assumptions remain unaltered in the dominant system. Thus, the service rate of each node $k$ is given by:

$$\mu_s = \Delta_k \overline{P}_{k2,sd} \frac{K \lambda_{es}}{1 - \lambda_{ep}(\lambda_p / \mu_p)} P_I a. \tag{3.33}$$

From (3.25), a packet is served from the relay queue with probability $\bar{a}$ if the SU cluster queues are non-empty or with a probability one if the SU cluster queues are empty. The probability of the SU cluster being empty can be expressed as the case at which all the SU queues are empty, $P_{(Q_s=0)} = (1 - \frac{\lambda_s}{\mu_s})^K$. Consequently, $P_{(Q_{ps} \neq 0)} = 1 - (1 - \frac{\lambda_s}{\mu_s})^K$. It can be verified that:

$$\mu_{ps} = \overline{P}_{k1,sd} \frac{K \lambda_{es}}{1 - \lambda_{ep}(\lambda_p / \mu_p)} \left\{ \bar{a} + a \left(1 - \frac{\lambda_s}{\mu_s}\right)^K \right\} P_I. \tag{3.34}$$

Substituting (3.33) in (3.34),

$$\mu_{ps} = \overline{P}_{k1,sd} \frac{K \lambda_{es}}{1 - \lambda_{ep}(\lambda_p / \mu_p)} P_I \left\{ \bar{a} + a \left(1 - \frac{\lambda_s}{\Delta_k \overline{P}_{k2,sd} \frac{K \lambda_{es}}{1 - \lambda_{ep}(\lambda_p / \mu_p)} P_I a}\right)^K \right\}. \tag{3.35}$$

Unlike (3.31), $\mu_{ps}$ is function of service policy $a$ and for $K = 1$, service policy $a$ will vanish, and consequently $\mu_{ps}$ will be independent on $a$ as in (3.12). For the SU relay queue stability, with $\mu_{ps}$ derived in (3.16), it can be showed that,

$$\lambda_{ps} < \mu_{ps},$$

$$\lambda_p < \mu_p \frac{\overline{P}_{k1,sd}}{P_{ps,pd}\overline{P}_{ps,k*}\lambda_{ep}} \frac{K\lambda_{es}}{1-\lambda_{ep}(\lambda_p/\mu_p)} P_I \left\{ \overline{a} + a\left(1 - \frac{\lambda_s}{\Delta_k \overline{P}_{k2,sd}\frac{K\lambda_{es}}{1-\lambda_{ep}(\lambda_p/\mu_p)}P_I a}\right)^K \right\}. \quad (3.36)$$

From (3.1), (3.34), (3.36), the stability region of system (*II*) would be bounded by:

$$R2 = \left\{ \begin{array}{c} (\lambda_p, \lambda_s): \\ \lambda_s < \Delta_k \overline{P}_{k2,sd}\frac{K\lambda_{es}}{1-\lambda_{ep}\left(\frac{\lambda_p}{\mu_p}\right)} P_I a, \\ \lambda_p < \mu_p \frac{\overline{P}_{k1,sd}}{P_{ps,pd}\overline{P}_{ps,k*}\lambda_{ep}} \frac{K\lambda_{es}}{1-\lambda_{ep}\left(\frac{\lambda_p}{\mu_p}\right)} P_I \left\{ \overline{a} + a\left(1 - \frac{\lambda_s}{\Delta_k \overline{P}_{k2,sd}\frac{K\lambda_{es}}{1-\lambda_{ep}\left(\frac{\lambda_p}{\mu_p}\right)}P_I a}\right)^K \right\} \end{array} \right\}$$

(3.37)

In Fig. 3.15, we show the stability region at three values of service policy $a$ of the dominant system (*II*). As showed in (3.34), the maximum sustainable ($\lambda_s$) increases as $a$ increases when the SU tends to serve its own nodes at the expense of the relay queue. For small $\lambda_s$ values, the relation between $\lambda_s$ and $\lambda_p$ forms a quadratic function for $K = 2$ as shown in (3.36). As $\lambda_s$ increases to certain limit, there is no real values of $\lambda_p$ satisfies this quadratic equation. So this limit of $\lambda_s$ is the maximum allowed arrival rate at each SU node complies with the stability constraint of $Q_{ps}$.

The maximum allowable PU arrival rate is from (3.23), $\mu_p = \lambda_{ep}(\overline{P}_{ps,pd} + P_{ps,pd}(1 - [P_{ps,k}]^K)) = 0.38$. From (3.36), with $K = 2$, the quadratic function between $\lambda_p$ and $\lambda_s$ is as:

$$\left(\lambda_{ep}\frac{\lambda_p}{\mu_p}\right)^2 \left[-P_{ps,pd}\overline{P}_{ps,k*} - P_{(Q_{ek}\neq 0)}\overline{P}_{k1,pd}\right]$$

$$+\left(\lambda_{ep}\frac{\lambda_p}{\mu_p}\right)\left[P_{ps,pd}\overline{P}_{ps,k*} + 2K\lambda_s + 2P_{(Q_{ek}\neq 0)}\overline{P}_{k1,pd}\right]$$

$$-\left[P_{(Q_{ek}\neq 0)}\overline{P}_{k1,pd} - 2K\lambda_s + \frac{(K\lambda_s)^2}{aP_{(Q_{ek}\neq 0)}\overline{P}_{k2,sd}}\right] < 0. \qquad (3.38)$$

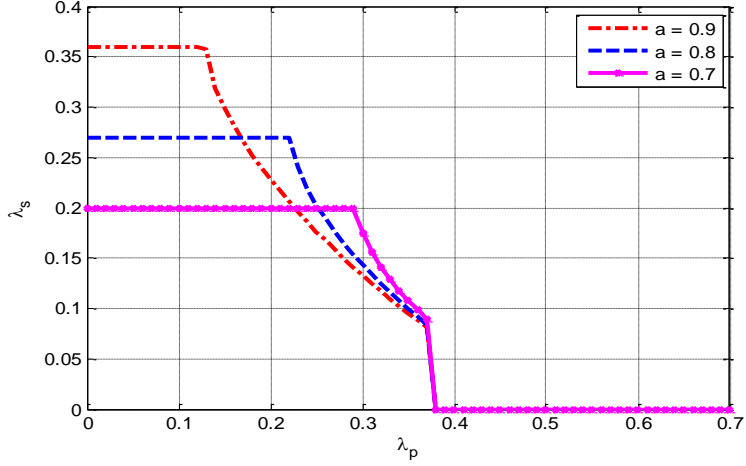

**Figure 3.15** Stable throughput region for dominant system (*II*) at different values of service probability $a$ ($\lambda_{ep} = 0.5, \lambda_{es} = 0.5$)

### 3.3.5 Performance Analysis

In this section, we obtain the stability region for overall cooperative system with energy harvesting as the union of the two dominant systems, $R = R_1 \cup R_2$. We let $K = 2$, $\overline{P}_{ps,pd} = 0.3$, $\overline{P}_{ps,k} = 0.4$, $\overline{P}_{k,pd} = 0.8$, and $\overline{P}_{k,sd} = 0.8$. In Fig. 3.16, we show the stability region at three values of service policy $a$, $0.7, 0.8$, and $0.8$.

The boundaries of the stability region are from (3.32) and (3.37). For $K = 2$, the maximum sustainable $\lambda_s$ is from (3.31), $\Delta_k (1 - [P_{k,sd}]^2) = 0.47$ and the maximum sustainable $\lambda_p$ is from (3.23), $\mu_p = \lambda_{ep}\overline{(P}_{ps,pd} + P_{ps,pd}(1 - [P_{ps,k}]^2) = 0.38$. From (3.27), with $K = 2$, $P_{(Q_{ek}\neq 0)} = 1$ for $\lambda_{es} \geq 0.5$. Therefore, SU cluster with energy harvesting rate $\lambda_{es} \geq 0.5$ at each node is identical to the case without energy constraint on the SU cluster.

It is noted that for the same $\lambda_s$ value, the system with lower service probability $a$ can sustain higher PU arrival rate, $\lambda_p$. The stability region of the overall system over the union of service probability $a$ from zero to one is plotted in Fig. 3.17. The same maximum achievable $\lambda_s$ and $\lambda_p$ rates as in Fig. 3.16

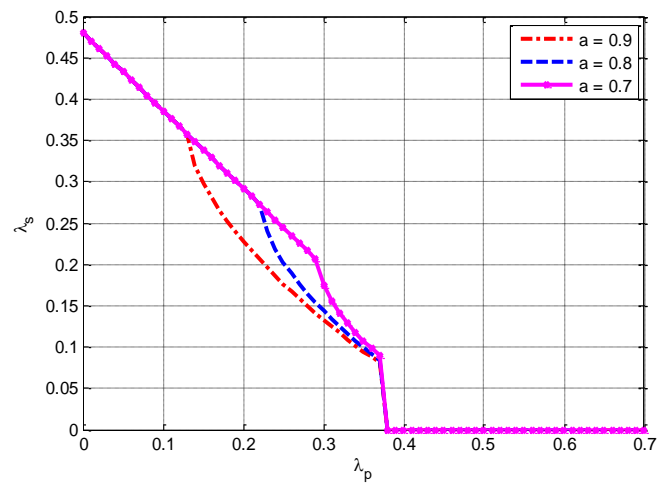

**Figure 3.16    Stable throughput region of the overall system at different service probability $a$ ($\lambda_{ep} = 0.5, \lambda_{es} = 0.5$)**

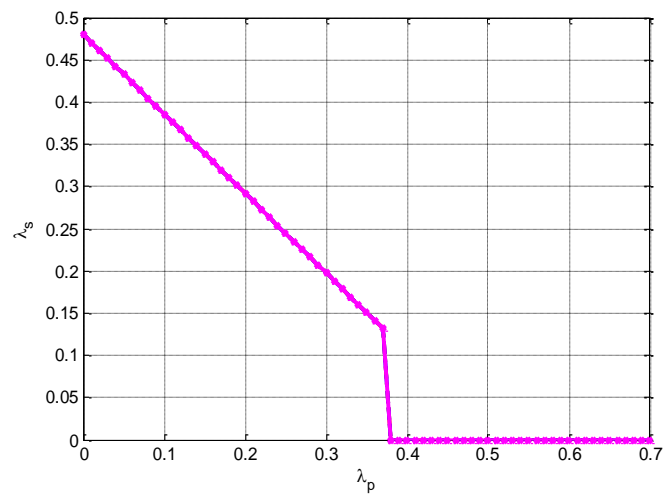

**Figure 3.17    Stable throughput region of the overall system ($\lambda_{ep} = 0.5, \lambda_{es} = 0.5$)**

## 3.3.6 Non-cooperative Energy Harvesting System

Here, we introduce the comparison between two systems, cooperative energy harvesting and non-cooperative systems under different energy harvesting rates. In addition, we depict the sufficient conditions for beneficial cooperation in case of SU cluster structure. For the PU in non-cooperative system, the same as the previous single SU system and $\mu_p$ as derived in (3.19).

Also, a packet is served from any SU node $k$ when the PU is idle,

$$\mu_s = \overline{P}_{k2,sd}\, \Delta_k \cdot \frac{K\,\lambda_{es}}{1 - \lambda_{ep}\,(\lambda_p/\mu_p)} P_I. \tag{3.39}$$

The stability regions of the cooperative and non-cooperation multiuser systems without energy constraint are plotted in Fig. 3.18(a), for $K = 2$ and the same system parameters as before. The maximum sustainable SU throughput is $(\overline{P}_{k2,sd}\, \Delta_k)$ for the two systems. It is shown that the maximum sustainable PU arrival rate is larger in the cooperative system. Also, cooperative system provides better SU sustainable arrival rate for all possible values of $\lambda_p$. Therefore, cooperation between PU and SU is beneficial for both users under the mentioned channel parameters.

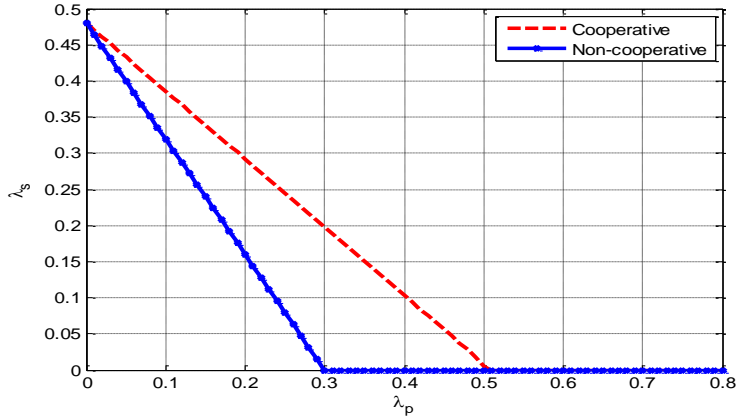

(a)

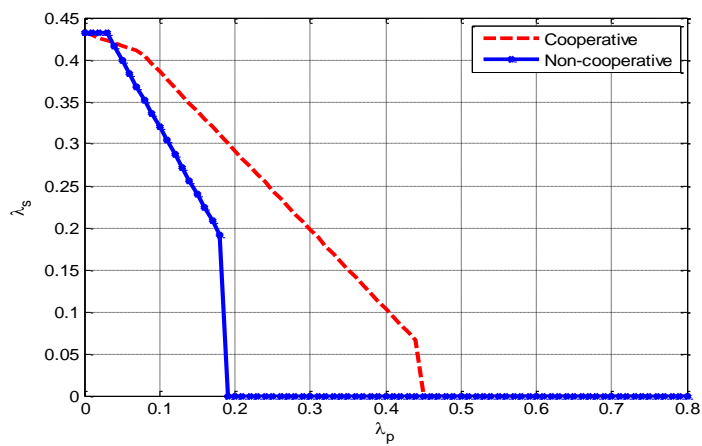

(b)

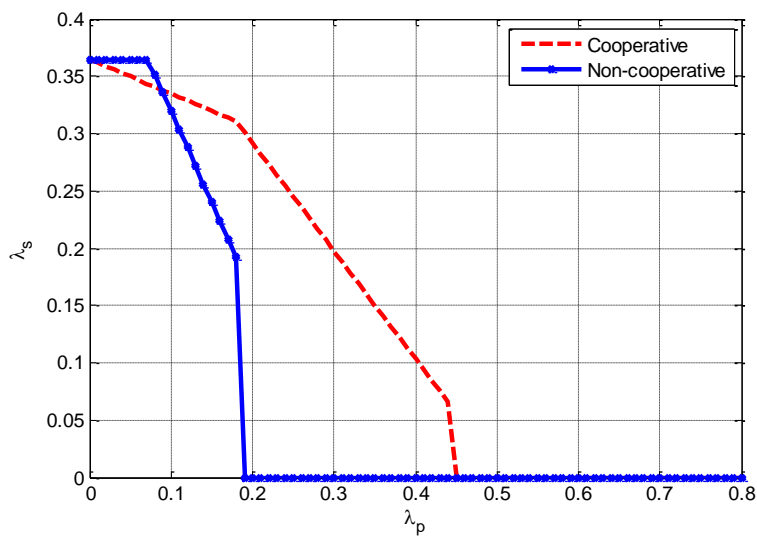

(c)

**Figure 3.18　Stable throughput region of the cooperative and non-cooperative systems $K = 2$ (a) $(\lambda_{ep} = 1, \lambda_{es} = 1)$ (b) $(\lambda_{ep} = 0.5, \lambda_{es} = 0.45)$ and (c) $(\lambda_{ep} = 0.5, \lambda_{es} = 0.38)$**

Taking into account the effect of energy harvesting, $\lambda_{ep}$ and $\lambda_{es}$ are considered less than unity. Fig. 3.18(b) and Fig. 3.18(c) consider the effect of energy constraints with $\lambda_{ep} = 0.3$, $\lambda_{es} = 0.45$ and $\lambda_{ep} = 0.3$, $\lambda_{es} = 0.38$, respectively. Figure 3.18(b) and Fig. 3.18(c) show that non-cooperative energy harvesting system outperforms the cooperative energy harvesting system for low values of $\lambda_p$ in terms of SU throughput. As $\lambda_p$ increases, the two systems intersect providing the same SU allowable throughput, after which, the cooperative system is the prevalent one. Also, it is noticed that this value of intersection is lower for higher $\lambda_{es}$.

The value of $\lambda_p$ of intersection, $\lambda_{pi}$, can be calculated as follows, we would equate $\mu_s$ from (3.31) with $\mu_s$ from (3.39) and rearranging:

$$\lambda_{pi} = \frac{1 - K \lambda_{es}}{\Omega}, \tag{3.40}$$

where $\Omega$ is a constant function of the channel outages probabilities

$$\Omega = \frac{1}{\overline{P}_{ps,pd}} - \frac{P_{ps,pd}\overline{P}_{ps,k*}}{\overline{P}_{k1,sd}(\overline{P}_{ps,pd} + P_{ps,pd}\overline{P}_{ps,k*})}, \tag{3.41}$$

Equation (3.40) shows that: $\lambda_{pi}$ is a function of the system channel parameters, SU cluster size, and SU energy harvesting rates. In addition, the value of $\lambda_{pi}$ decreases as $\lambda_{es}$ increases. With the aforementioned values of $\lambda_{es}$, $\lambda_{pi} = 0.03$ for Fig. 3.18(b) and $\lambda_{pi} = 0.08$ for Fig. 3.18(c). This can be emphasized as follows; the SU cluster with higher energy levels at its nodes is supposed to has a better opportunity to cooperate with the PU and leverage the better channel conditions. Also as $K$ increases, the channel diversities between PU source, the SU cluster and SU cluster, PU destination increase that makes the cooperative system outperforms the non-cooperative system even for low $\lambda_p$ values.

To investigate the effect of SU cluster size on the PU performance, we plot the value of $\lambda_{pi}$ and the maximum PU sustainable arrival rate, $\lambda_{pm}$ versus the SU cluster size in Fig. 3.19 and Fig. 3.20, respectively. Parameters used are: $\lambda_{ep} = 0.5$, $\lambda_{es} = 0.2$, $\overline{P}_{ps,pd} = 0.3$, $\overline{P}_{ps,k} = 0.4$, and $\overline{P}_{k,pd} = 0.6$. Examining Fig. 3.19, it is intuitive from (3.40) that $\lambda_{pi}$ is monotonically decreasing as $K$ increases because of the higher diversity SU cluster provides. At small $K$, the value of $\lambda_{pi}$ decreases as $\overline{P}_{k,sd}$ increases since the SUs gain more benefits from the cooperation when the channels to the common destination is better. As $K$ increases, $\lambda_{es} = 0.2$ will eventually equal $1/K$ and $\lambda_{pi} = 0$ from (3.40).

This leads that the cooperative system outperforms the non-cooperative system for all PU arrival rate.

Figure 3.20 shows that the maximum PU sustainable arrival rate, $\lambda_{pm}$, is increasing with the power of $K$ from (3.23) with higher diversities the SU cluster sustain for the PU packets. Also, the better PU-SUs channel, $\overline{P}_{ps,k}$, the higher of $\lambda_{pm}$. All these values of $\lambda_{pm}$ converge to the same value of $P_{(Q_{ep} \neq 0)} = 0.5$ at $K$ approaches infinity.

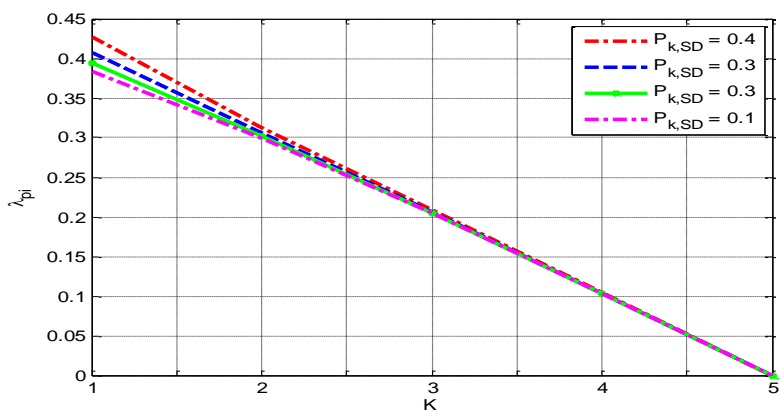

**Figure 3.19　　$\lambda_{pi}$ v.s. SU cluster size ($K$)**

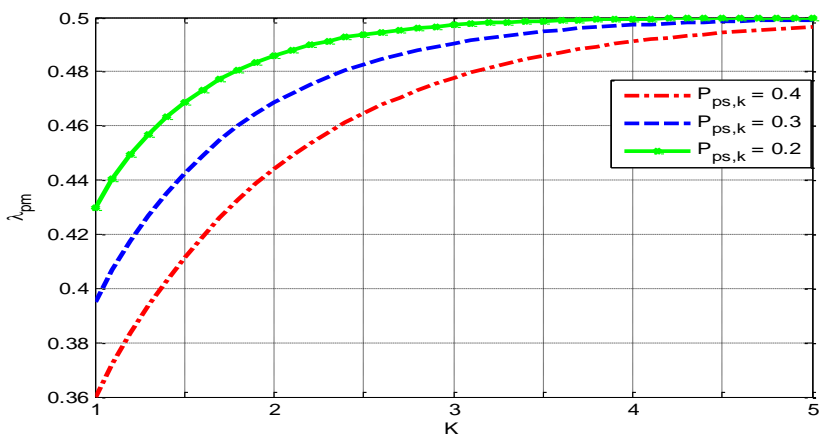

**Figure 3.20　　$\lambda_{pm}$ v.s. SU cluster size ($K$)**

# CHAPTER FOUR
# COOPERATION AND UNDERLAY MODE SELECTION IN COGNITIVE RADIO NETWORK

## 4.1 INTRODUCTION

In this chapter, we propose we propose a technique for cooperation and underlay mode selection in cognitive radio networks. Hybrid spectrum sharing is assumed where the SU can access the PU channel in two modes, underlay mode or cooperative mode with admission control. Typically, cooperation possesses better performance than non-cooperation since the exploitation of the better channel conditions through the secondary network. Overlay spectrum sharing allows the SU to occupy the spectrum only when the PU is idle. Cooperation (collaboration) occurs when the SU admits the PU's packet into a relay buffer to transmit in the subsequent timeslot, which happens in timeslots when no successful transmission from PU source to the destination. In addition to accessing the channel in the overlay mode, secondary user is allowed to access the channel in the underlay mode by occupying the channel currently occupied by the primary user but with small transmission power. Adding the underlay access mode provides more opportunities to the secondary user to transmit its data. It is assumed that the secondary user can only exploit the underlay access when the channel of the primary user is at good conditions, or predicted to be in non-outage state. Therefore, the secondary user could switch between underlay spectrum sharing and cooperation with the primary user according to the state of the primary channel.

The destination sends a feedback message with the received SNR (channel quality status) to the PU source, which is observed by the SU. The SU exploits the SNR feedback messages to build and update a belief function about the primary channel transmission quality. The secondary user exploits its observation and belief function about the primary channel to take the appropriate decision each timeslot. If the channel sustains good transmission conditions, the secondary user can access in the underlay mode. On the other hand, if primary channel is predicted to be in outage, the secondary user not only becomes silent, but also admits the packet in the relay queue to cooperate with the primary user in case of failure in the direct transmission. Our system depends on primary user channel quality and the *action-reward model* of the system to choose the appropriate decision each timeslot, whether to cooperate with the PU or to access the channel in the underlay mode.

Results reveal that the proposed model attains noticeable improvement in the system performance in terms of maximum secondary user throughput than the conventional cooperation and non-cooperation schemes. To the best of our knowledge, hybrid spectrum sharing depending on the primary channel has not been studied before.

## 4.2 RELATED WORKS

Recently, partial cooperation between the SU and PU has gained a lot of attention in cognitive radio research. Specifically, SU admits some of the undelivered packets to relay these to the primary destination. Fine-tuning on the cooperation between the SU and PU is achieved by adding admission control at the relay queue. This controls the behavior of the network and provides adaptable enhancement to any of the PU and SU.

In [43], authors studied the queues stability and delay in a cooperative cognitive radio networks. A system of one PU and one energy-limited SU is studied. The SU has two data queues and a battery for energy storage. A PU's packet is admitted to be stored in the relay queue with an admission probability. In addition, the SU serves either of the two data queues with certain service probabilities. The system throughput and delay are proved to change with varying the service and admission probabilities and by the finiteness of energy. Authors resorted to the dominant system approach for the system analysis and characterization of the stability region. The moment generating function approach is employed in the derivation of the average delay encountered by the packets of both PU and SU. The stability region and the average packet delay of two systems, with and without energy constraints, are compared.

Results reveal that, the maximum achievable PU arrival rate is non-decreasing function of acceptance ratio at the relay queue in the non-energy-constrained system. The maximum achievable PU arrival rate is non-decreasing function of acceptance ratio in the energy-constrained system as the SU battery is able to serve the excess PU relayed packets. The delay encountered by the PU packets is shown to be decreasing function of PU arrival rate with higher delay found in the energy-constrained system.

A network consists of a single cognitive radio transmitter–receiver pair shares the spectrum with two primary users is analyzed in [44]. Each primary user has a data traffic queue, whereas the cognitive user has three queues; one storing its own data queue while the other two are the relaying queues corresponding to the relayed packets from the two PUs. A cooperative cognitive MAC protocol for the proposed network is suggested, where the SU exploits the idle periods of the two PU bands. Traffic arrival at each relay queue is controlled via a tunable admittance factor, while relay queues service is controlled via channel access probabilities assigned to each queue based on the band of operation.

The stability region of the proposed protocol is plotted with focus on its maximum expected throughput. The performance gains of the proposed cooperative cognitive protocol with admission control are shown to outperform the non-cooperative and conventional cooperative systems. The SU may combine/merge the available primary orthogonal bands to increase the probability of successful packet reception, which in turn increases its service rate.

In contrast to [44], authors in [45] considered a cognitive radio system with two SUs who can cooperate with single PU. In addition to its own data queue, each secondary user has a queue to store the PU packets that are erroneously decoded by the PU receiver. Secondary users accept the relayed packets with a certain probability and transmit randomly from either of their queues whenever they are both nonempty. Admission and service probabilities are optimized to expand the maximum stable throughput region of the system.

One SU, second ranked, senses the channel to detect the possible activity of the PU and the other SU, first-ranked. It transmits, if possible, starting after the sensing time from the beginning of each timeslot. The second-ranked SU increases its transmission rate so that it can transmit one packet in one timeslot but with higher probability of outage than the first-ranked SU. It is shown the potential advantage of this ordered system over the conventional random access system.

The maximum throughput of a rechargeable SU sharing the spectrum with a PU is characterized in [46]. The SU harvests its energy packets (tokens) from the environment with a certain harvesting rate. The harvested energy at the SU is stored at a buffer. A new cooperative cognitive relaying protocol is proposed that allows the SU to relay a fraction of the undelivered PU packets by introducing admission control at the relaying queue. A channel with Multi-Packet Reception (MPR) model is assumed, where concurrent transmissions can survive from interference with certain probability.

This proposed protocol exploits the PU queue burstiness and receivers' MPR capability. Numerical results showed the benefits of cooperation, receivers' MPR capability, and SU energy harvesting rate on the system performance. Authors provided two inner bounds and one outer bound on the secondary throughput, and showed that the bounds are coinciding when the secondary energy queue is always backlogged (no energy constraint at the SU).

In [47], a time-slotted system model is analyzed where the nodes are randomly accessing a common receiver. A network of three nodes denoted as source, relay, and destination is presented. The relay may receive some of the packets transmitted from S and then relay those packets to the common destination. The relay node has two queues for the exogenous packet arrivals and the endogenous packet arrivals from the source.

Authors did not pay attention to the order by which the packets are served from each queue. Instead, it is assumed that the relay maintains a single queue and merges all the arrivals into a single queue, as the achievable stable throughput region is not affected.

The relay node is equipped with a flow controller. Objective is to find the optimum value of acceptance ratio that maximizes the stability region of the overall system. This value represents the cooperation degree between the two nodes that maximizes the stability region. The stability region of the overall system is plotted and results reveal that, region of partial cooperation (cooperation with admission control) contains the regions of the other cases specifically, non-cooperative and full cooperative systems. Therefore, the partial cooperation scheme is superior compared to the rest schemes.

Authors studied the problem of cooperative cognitive radio systems in [48]. Secondary user has limited relaying queue for the overheard primary packets. The stable throughput region of a cognitive radio network with a finite relaying buffer at the secondary user is characterized. Objective was to maximize the secondary user throughput while guaranteeing the stability of the primary user queue. The packet admission and queue selection for serving at the secondary user are dependent on the state of the finite relaying buffer.

The packet admission probability in the system depends on the number of packets in the relay queue. If the SU admits the packet and store it in the relay buffer, it sends back an ACK so that the PU drops the packet from its queue. The queue selection to be served in the system also depends on number of packets in relay queue. Regarding the finite queue size of relay buffer, it should considered for the blocked packets when the relaying queue is full at certain timeslot; i.e. when it filled with K packets. This scenario is handled by forcing the system to reject any relayed packets from the PU.

This system is assumed non-work-conserving because there are possible cases where the system might waste timeslots despite having packets to be delivered. Despite of the non-work-conserving property, the SU tends to serve its own data queue with probability one whenever the relay queue is empty.

Sequential decision-making has been employed in many works of cognitive radio research to model a framework of the system behavior. In [49], the problem of opportunistic spectrum access in Rayleigh fading channel is studied. Channels are characterized by both channel quality and the probability of being occupied by primary users. First, a finite-state Markov channel model is implemented to represent a fading channel. Second, by joint probing each channel quality and activities of PUs, a two-dimensional partially observable Markov decision process framework is suggested for OSA. Due to the high complexity of such a problem solution, a greedy strategy is proposed where a SU selects a channel that has the best-expected data transmission rate to maximize the immediate reward. When compare with the optimal strategy that considers both immediate and future rewards, the greedy strategy brings low complexity in design and almost ideal performance.

The integration of spectrum sensing function and dynamic spectrum access scheme is introduced in [50] to satisfy the primary users' interference tolerance. In this work, authors proposed a Hybrid Non-Cooperative MAC (HNC-MAC) protocol for the SU to consider the spectrum sensing function and dynamic spectrum selection scheme. Using HNC-MAC protocol, SU can get the maximum throughput under PU's collision constraint and perform vacant channel selection with POMDP algorithm. The timeslot structure in HNC-MAC protocol is composed of four parts; idle channel selection, channel sensing, data transmission and acknowledgment.

At the beginning of each timeslot, based on the history statistics of PU's activity over the N channels, SU using HNC-MAC protocol can probe an idle channel as the target access channel. Then, the sensing of selected channel is performed to reduce the collision probability with PU since missed detection is assumed. When the SU senses an idle channel, the SU's packet data could be transmitted. A the end of each timeslot, the transmission is over, an ACK from SU's destination is sent back to the SU source to inform when the packet is correctly decoded to synchronize the channel selection upon next slot. Simulation results show that the HNC-MAC protocol effectively improves the SU's throughput.

Authors studied a hybrid underlay-overlay cognitive radio with energy harvesting in [51]. The secondary user can harvest energy from the primary user's signal as well as from the other ambient sources e.g. wind and vibration. The harvested Energy from the primary user's signal is available only when the primary channel is active.

The secondary user is allowed to operate in one of three transmission modes; overlay and underlay to maximize its throughput, or remains in the sleep mode in order to conserve its harvested energy, or harvests energy from the primary transmission in order to maximize the remaining harvested energy.

Results showed that this system provides 60% improved throughput than overlay-only cognitive radio and 43% enhanced throughput than a hybrid cognitive radio system harvesting energy only from the ambient sources.

## 4.3 SYSTEM MODEL

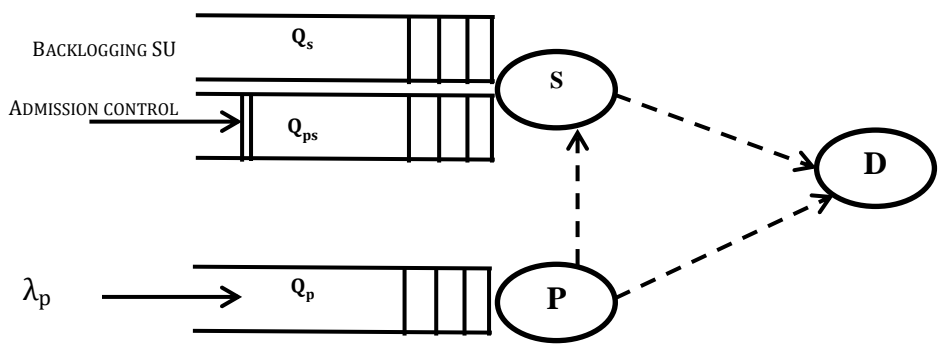

**Figure 4.1    System model**

Fig. 4.1 depicts our system model. The system is composed of one PU and SU transmitters send their data to a common destination. It is supposed that in a time-varying environment both users accessing a licensed fading channel with bandwidth B Hz. Secondary user have to carry out spectrum sensing to detect spectrum occupancy each timeslot. Primary user has one data queue, $Q_p$. The arrival process at $Q_p$ is modeled as Bernoulli arrival process with mean $\lambda_p$ [packets/slot]. Considering the SU, it is represented by two data queues, $Q_s$ and $Q_{ps}$. $Q_s$ is an infinite capacity buffer for storing the SU's own packets which is assumed to be backlogged. $Q_{ps}$ is a buffer for storing the admitted primary packets for relaying, that have not been successfully received directly by the destination.

Primary user transmits packets from $Q_p$ whenever the queue is non-empty. Primary user removes the transmitted packet from its queue when it is followed by an ACK message from the destination or from the SU. The destination sends a feedback message to the PU source to inform whether the transmission was successful or not. The feedback message contains two fields; one field represents ACK/NACK types, and another field represents the SNR level received by the destination. If the received SNR level at the PU destination was above certain threshold, the PU source receives a positive ACK message corresponding to successful transmission, and vice versa. Consequently, the channel quality status of the primary link is known from the received SNR at the destination.

Due to the broadcast nature of wireless transmission, the SU can overhear the ACK/NACK messages from the PU destination to the PU source, it can send independent ACK feedback message if the PU source-destination link is in outage and the PU packet is correctly received and admitted at the SU source. Additionally, the SU exploits the SNR feedback associated with the ACK/ NACK message to build and update a belief function that is used in conjunction with the reward function to decide the action (access mode) in each timeslot. The secondary user exploits its observation about the primary channel to take the appropriate decision each timeslot. If the channel sustains good transmission conditions, the secondary user can access in the underlay mode. On the other hand, if primary channel is predicted to be in outage, the secondary user not only becomes silent, but also admits the packet in the relay queue to cooperate with the primary user.

Our proposed hybrid spectrum sharing model works as follows: each timeslot the PU channel is sensed. If the PU is sensed to be idle, the SU accesses the channel to serve a packet from $Q_{ps}$ if nonempty otherwise it serves its own data queue. If the channel is sensed as busy, the SU has two decisions (actions) corresponding to two operation modes. In the underlay mode, and depending on the belief function about the channel between the PU source and the destination, the SU sends its data concurrently with the PU transmission with a limited power. In the cooperative mode, the SU chooses to accept the PU packet and inserts it in the relay queue. In the time slots when the channel between the PU source and the destination is in outage, if the SU received the PU packet correctly, the packet will be stored in the relay queue and the SU will bear the responsibility to deliver this packet by sending ACK message to the PU source.

The assumed model of the PU channel is tenable to build the belief function and use it to predict the future of the channel quality, as discussed in next subsections. Furthermore, the SU has a reward function of his decision; this reward is defined as the profit gained by the SU corresponding to the chosen action. We further propose a flowchart of our hybrid spectrum-sharing model, which is shown in Fig. 4.2.

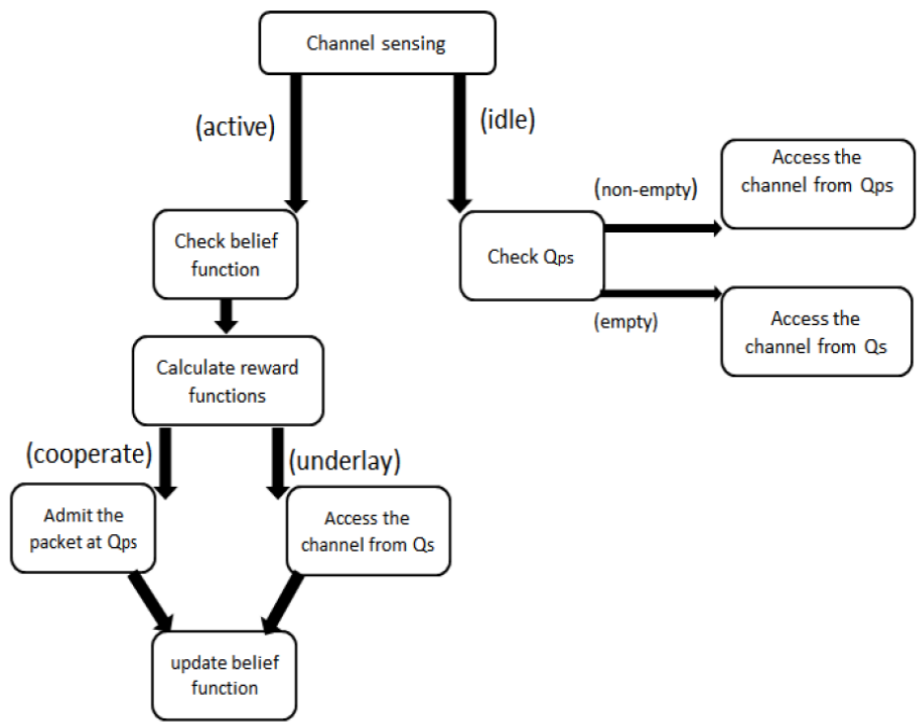

Figure 4.2    Flow chart of hybrid spectrum sharing

### 4.3.1 Channel Models

In this subsection, we explain the channel model of the link between the PU source and the destination. All channel models in the network are assumed to have the same model. A fading channel in a time-varying environment is modeled as Finite State Markov Chain FSMC model (Fig. 4.3).

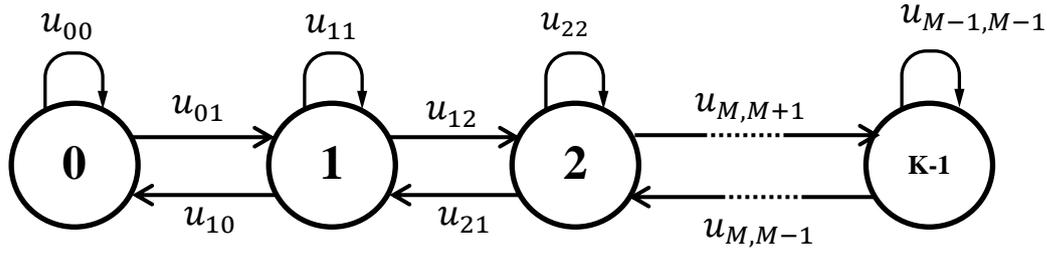

**Figure 4.3**  FSMC channel model

This model is introduced to characterize the channel quality levels of the fading channel, these levels are given as:

$$\Gamma_m = e^{m\eta/B} - 1 \qquad m \in \{0, 1, \cdots, M-1\}, \qquad (4.1)$$

where $0 = \Gamma_0 < \Gamma_1 < \cdots \Gamma_{m-2} < \Gamma_{m-1} < \infty$ and $\eta$ represents the different between adjacent SNR levels in terms of achievable data rate, in units of bps. The source transmits a training sequence in each packet and in turn, the destination receives them and sends feedback to the transmitter with the received SNR [52]. Accordingly, the received instantaneous SNR $\gamma$ is partitioned into $M$ non-overlapping levels. If $\Gamma_m < \gamma < \Gamma_{m+1}$, the current SNR state of the channel is regarded as $m$.

The packet transmission is assumed to last for a whole timeslot, and the length of the timeslot is large enough to make the assumption valid, where the SNR states change according to the FSMC from certain state only to the adjacent states or regain the same level. In the next subsections, we are going to derive an expression for steady state distribution of each level of the quantized channel levels and the rate of crossing each level. These parameters will be used to calculate state transition probabilities between channel quality levels. In section 4.3.2, we will adopt state transition probabilities to build and update the belief function at the SU.

In such a multipath propagation environment with fading channel, the received SNR is a random variable obeys the exponential distribution with the Probability Density

Function (PDF) $= p(\gamma) = \frac{1}{\gamma_0} e^{-\gamma/\gamma_0}$ and **steady state distribution** given by $\pi_m$ of the level $m \in \{0, 1, \cdots, M - 1\}$.

$$\pi_m = \int_{\Gamma_m}^{\Gamma_{m+1}} p(\gamma) \, d\gamma$$

$$\boldsymbol{\pi_m} = e^{-\Gamma_m/\gamma_0} - e^{-\Gamma_{m+1}/\gamma_0} \quad m \in \{0, 1, \cdots, M - 1\}, \quad (4.2)$$

and the **level crossing rate** function is $\Lambda(m)$ for a given level $m \in \{0, 1, \cdots, M - 1\}$,

$$\Lambda(m) = \sqrt{\frac{2\pi \, \Gamma_m}{\gamma_0}} \, f_{dopp} \, e^{-\Gamma_m/\gamma_0} \qquad m \in \{0, 1, \cdots, M - 1\} \quad (4.3)$$

where the level crossing rate function $\Lambda(m)$ denotes the average number of times per time interval that the fading signal crosses a certain signal level [53], and $f_{dopp}$ represents the Doppler frequency of the mobile terminal. It is involved here since it was found that the envelope of the fading levels of such a channel is shaped by the Doppler frequency.

The adjacent transfer method proposed in [53] is assumed for the transitions between SNR levels, which assumed that any transition only could happen between adjacent states or between the same states:

$$P_{m,l} = 0, \text{ if } |m - l| > 1, \qquad (m, l) \in \{0, 1, \cdots, M - 1\}. \quad (4.4)$$

The state transition probabilities can be obtained directly from the steady state distribution of each level $m \in \{0, 1, \cdots, M - 1\}$ and the level crossing rate function $\Lambda(m)$ of the corresponding level, Therefore, the state transition probabilities can be obtained as follows:

$$\begin{cases} u_{m,m+1} = \frac{\Lambda(\Gamma_{m+1})}{\pi_m} \tau_{pkt} & m = 0, 1, \cdots, M - 2 \\ \\ u_{m,m-1} = \frac{\Lambda(\Gamma_m)}{\pi_m} \tau_{pkt} & m = 0, 1, \cdots, M - 1 \end{cases} \quad (4.5)$$

where $\tau_{pkt}$ represents the packet duration time [52]. Since the summation of each transition probabilities in all states is one, the probability to transit to the same state is given by $u_{m,m}$

$$u_{m,m} = \begin{cases} 1 - u_{01} & m = 0; \\ 1 - u_{m,m+1} - u_{m,m-1} & m = 2,\cdots,M-2 \\ 1 - u_{M-1,M-2} & m = M-1 \end{cases} \quad (4.6)$$

The decision at timeslots when the PU is active is whether to accept the PU packet in the relay queue, or to access the channel concurrently in the underlay mode. This decision (action) is built upon the belief function at the SU. The SU builds and updates a belief function about the probability that the SNR level of the PU channel will be in certain SNR level among predefined levels.

In the next subsection, we introduce the belief function expression and the suggested reward function used to map the probabilities of the belief function to the chosen action. The decision is selected to maximize the current reward of a finite horizon problem

### 4.3.2 Belief and Reward Functions

As introduced in section 4.3.1, the SU maintain and update a belief vector representing sufficient statistical information about the PU channel, as the actual quality state is unknown. The SU can know the exact state of the PU channel only when the PU is active as the channel state could be overheard from the feedback message. Therefore, an important concept called quality vector is defined to describe the channel quality information and is defined as follows:

$$\Theta_n(t) = [\, \beta_n^0(t) \quad \beta_n^1(t) \quad \beta_n^2(t) \quad \ldots\ldots \quad \beta_n^m(t) \quad \beta_n^{M-1}(t) \,] \quad (4.7)$$

$$n \in \{\, 0, 1, 2, \quad , N \,\}$$

$$m \in \{0, 1, \cdots, M-1\}$$

where $\Theta_n(t)$ is the belief function about the PU channel quality state in timeslot n, and $\beta_n^m(t)$ is the probability that channel quality state in time index n lies in state $m$. It is clear that $\sum_m \beta_n^m(t) = 1$ and the probability $\beta_n^m(t) < 1$.

The update of the belief function each timeslot is dependent on whether the PU is active or not. If the PU is active, the SU listens to the feedback message and obtain the state of the channel quality in that timeslot. If the PU is idle, the update of the belief function is calculated by averaging the current belief over all transition probabilities.

$$\beta_n^l(t+1) = \begin{cases} u_{m,l} & \sigma(t) = feedback, \ level = m \\ \sum_m \beta_n^m(t) u_{m,l} & \sigma(t) = no\ feedback \end{cases} \quad (4.8)$$

where $\sigma(t) \in \{feedback, no\ feedback\}$. Considering the direct transmission between the primary source and the destination, the achievable rate can be calculated as:

$$\acute{\eta}_P^{DT} = \log_2\{1 + SNR\} \quad (4.9)$$

where SNR is the signal to noise ratio of the direct transmission between primary source and the destination, which is assumed before as exponential random variable with the assumed characteristics in section 4.5. The mean of the SNR random variable is $\gamma_{op}$. The target transmission rate of the PU link is $R_p$, which is constant and determined be the required QoS. The link of direct transmission is said to be non-outage if the target transmission rate of the PU is lower than the achievable rate.

$$P_{DT,no\ outage}^p = \Pr\{R_p < \acute{\eta}_P^{DT}\} \quad (4.10)$$

$$P_{DT,no\ outage}^p = \exp\left(-\frac{\rho_p}{\gamma_{op}}\right) \quad (4.11)$$

where $\rho_p = 2^{R_p} - 1$ and $P_{DT,outage}^p = 1 - \exp\left(-\frac{\rho_p}{\gamma_{op}}\right)$

for the link of direct transmission to be in non-outage state, the instantaneous SNR value at the desired destination should be greater than certain SNR threshold obtained from (4.9),

$$\Gamma_{threshold} = 2^{R_p} - 1 \quad (4.12)$$

On the first hand, the reward function is expressed as an increasing function when the SU tends to cooperate with the PU while the link of direct transmission is at bad transmission conditions. Therefore, the SU is going to admit the PU packets in the relay queue when the direct transmission cannot sustain the required target rate due to the low instantaneous SNR. On the other hand, the function is expressed as increasing function when the SU tends to choose the underlay access while the link of direct transmission is at good transmission conditions. For every state m $\epsilon$ M, we have two reward values corresponding to two actions made by the SU, whether to cooperate with PU or access the channel in the underlay mode. We define the reward function as follows:

$$R\left(m, a_{cooperate}\right) = \begin{cases} A_m & \text{SNR} < \Gamma_{threshold} \\ 0 & \text{SNR} > \Gamma_{threshold} \end{cases} \quad (4.13)$$

$$m \epsilon \{0, 1, 2, \ldots, M-1\}$$

$$R\left(m, a_{underlay}\right) = \begin{cases} 0 & \text{SNR} < \Gamma_{threshold} \\ B_m & \text{SNR} > \Gamma_{threshold} \end{cases} \quad (4.14)$$

$$m \epsilon \{0, 1, 2, \ldots, M-1\}$$

Since the actual reward function cannot be obtained at the beginning of each timeslot due to the randomness of SNR, and the actual state of the channel quality is unknown, we resort to the expected reward function $\rho\left(\Theta_n(t), a\right)$. The expected reward function is calculated by averaging the reward values over the probabilities that SNR lies in certain state giving the reward value.

$$\rho\left(\Theta_n(t), a\right) = \sum_m \beta_n^m(t) \, R(m, a) \quad (4.15)$$

The expected reward function is calculated for the two actions and the action with the highest reward is chosen.

## 4.4 PERFORMANCE ANALYSIS

In this section, a typical scenario is considered in which the channel is slowly fading and the environment is time varying. For validation of the obtained results, the mentioned parameters previously used in [9] are used here in order to investigate how much the proposed system can approach the previously published one. System performance is evaluated by using the MATLAB simulator whose parameters are listed in table 4.1. It is assumed that the power of the underlay transmission is 20 % of the overlay power. The reward parameters for the two possible actions are listed in table 4.2.

**Table 4.1 Simulation parameters**

| | |
|---|---|
| fc | 10 B |
| $v$ | 2 m/s |
| $\tau_{da}$ | 100 ms |
| $\eta$ | 3 Mbps |
| M | 8 |
| $\gamma_{op}$ | 15 dB |
| $\gamma_{os}$ | 45 dB |
| $\gamma_{ops}$ | 20 dB |
| $R_p, R_s$ | 3.5 bps/Hz |
| Number of iterations | 200000 |

**Table 4.1 Reward parameters**

| m | 1 | 2 | 3 | 4 | 5 | 6 | 7 | 8 |
|---|---|---|---|---|---|---|---|---|
| $A_m$ | 5 | 5 | 5 | 5 | 0 | 0 | 0 | 0 |
| $B_m$ | 0 | 0 | 0 | 0 | 6 | 7 | 8 | 9 |

In Fig. 4.4, we plot the SU throughput versus the primary user arrival rate for the non-cooperation, conventional cooperation, and proposed hybrid cooperation methods. The figure reveals that the SU throughput of the proposed method is higher than the SU throughput of the non-cooperation and conventional cooperation methods. On the first hand, at low $\lambda_p$, the SU throughput of the proposed hybrid cooperation outperforms the other two methods. This can be interpreted as follows: the SU exploits the belief function about the PU's channel to access the channel in the underlay mode when it predicts good channel conditions besides cooperation with the PU. On the other hand, when $\lambda_p$ becomes higher, the SU throughput of the proposed method is only the non-zero as the SU could access the channel only in the underlay mode.

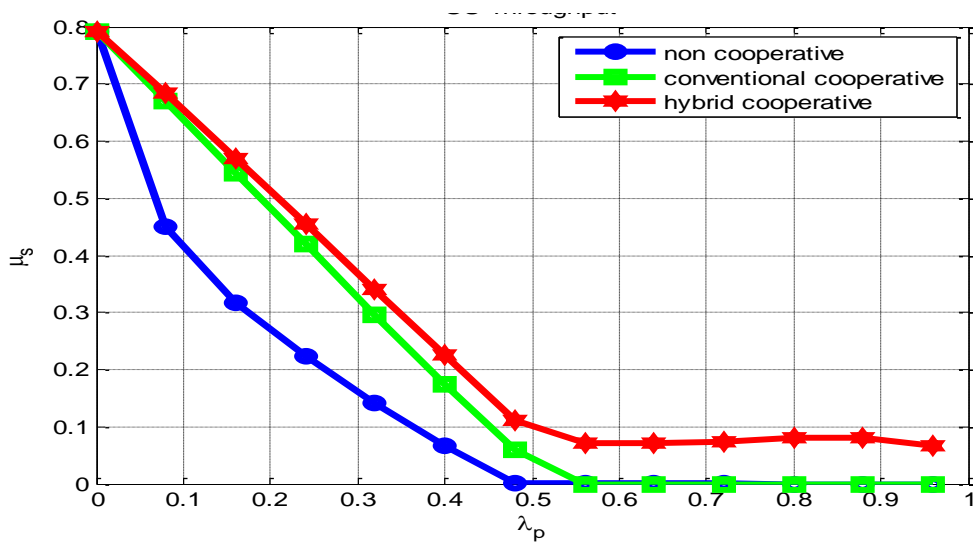

**Figure 4.4    SU throughput versus the PU arrival rate**

Figure 4.5 shows the maximum SU throughput for different underlay transmission power, 20%, 40%, and 60% of the overlay power. Results reveal that, increasing the underlay power of SU leads to lower SU throughput when $\lambda_p$ is small and higher SU throughput when $\lambda_p$ is high. This can be interpreted as, at low $\lambda_p$, the SU with higher underlay power causes more interference to the PU. The more interference at the PU results in more retransmission of PU undelivered packets either through the direct link or by the relaying SU. This leads to lower opportunities for the SU to access the channel in the overlay mode to serve his data. On contrary, the SU mainly accesses the channel in the underlay mode when $\lambda_p$ is high, so the higher underlay power gives the better SU throughput.

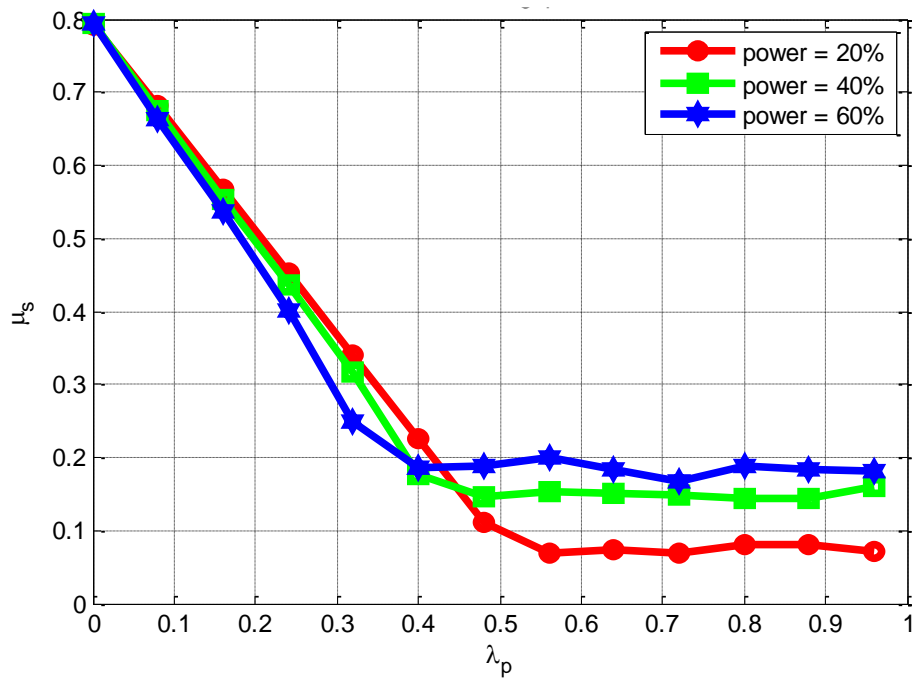

**Figure 4.5    SU throughput at different SU underlay power**

# CHAPTER FIVE
# SUMMARY AND CONCLUSIONS

## 5.1 CONCLUSION

In this thesis, we studied the topic of spectrum sharing in cognitive radio networks. A literature survey is involved focusing on cooperative spectrum sharing. The need for power supply in wireless nodes guided us to study energy harvesting in cognitive radio networks. Energy harvesting is found to be the solution of imposing reliable power supply in distributed wireless devices. In the first problem, we assumed a cooperative cognitive radio network of two scenarios. In the first scenario, a network of single primary and secondary users is studied. The secondary user is assumed to cooperate with the primary user to relay the undelivered packets in the direct path of primary transmission. In order to study the stability of this system, we represented mathematically the stability of each queue, and resorted to the dominant system approach to resolve such an interactive queuing system. System results showed the difference between the energy harvesting system and the system without energy constrains. Additionally, we characterized the difference between the cooperative and non-cooperative models with energy harvesting imposed on primary and secondary users.

In the second scenario, the system is extended to include a cluster of energy harvesting secondary users. The cluster is controlled via a cluster supervision block node to control the activities of each secondary user and the common relay queue. The enhancement to the primary user performance is shown to be function of the size of the secondary cluster.

In the second problem, we proposed a system in which the secondary user is allowed for the hybrid channel access. We studied in elaborate detail different allowable access modes for the secondary user. The secondary user is assumed to access the channel in two modes: either underlay mode or overlay mode with cooperation. The secondary user can switch to the appropriate mode according to the status of the primary channel. If the channel is estimated to be in good transmission conditions, the secondary user can access the channel concurrently with the primary user. On contrary, if the channel is assumed to be in outage, the secondary user cooperates with the primary user to deliver its packets to the common destination.

The system model and channel analysis are illustrated. Results reveal that the hybrid access outperforms the conventional cooperation and the non-cooperation access schemes. The system is simulated for different secondary user's power and different reward parameters to select the appropriate decision whether to cooperate with the primary user or not.

# PUBLICATION

# REFRENCES

# ملخص الرسالة

تقترح هذه الرسالة نموذجين جديدين من نماذج التعاون في شبكات الراديو الإدراكي. حيث يعتمد النموذجان علي وجود تعاون بين نوعين من المستخدمين لشبكات الراديو الإدراكي بحيث يتم الإستفادة من بعض قنوات الاتصال ذات الحالة الجيدة عند أحد المستخدمين لنقل المعلومات للمستخدم الاخر.

يهدف النموذج الأول المقترح فى الرسالة إلى إمداد مستخدمي الراديو الإدراكي بخلايا شحن الطاقة يمكنها إعادة الشحن أثناء التشغيل. تمثل خلايا شحن القدرة الكهربية مقترح لتشغيل المستخدم اللاسلكي لشبكات الراديو الإدراكي في حالة صعوبة التزويد بمصدر ثابت للقدرة الكهربية. تم إقتراح النموذج الجديد من أجل الإستغلال الأمثل للطاقة. تم إثبات أن هذا النموذج يمكن الحصول منه على أداء مقارب من شبكات الراديو الإدراكي التي تعمل بمصدر ثابت للقدرة الكهربية وفي بعض الظروف يكون الأداء متطابق في حالتي العمل بمصدر ثابت للقدرة الكهربية والعمل بخلايا لشحن الطاقة. تمت دراسة النموذج في حالة وجود مستخدم ثانوي وحيد وفى حالة وجود العديد من المستخدمين الثانويين.

يهدف النموذج الثاني المقترح فى الرسالة إلى إضافة خوارزم جديد لشبكات التعاون في استخدام الراديو الإدراكي. حيث يمكن التحكم فى درجة التعاون بين المستخدم الثانوي والمستخدم الأساسي لشبكات الراديو الادراكي. يتم العمل فى إتجاه زيادة التعاون بين المستخدم الثانوي والمستخدم الأول فى حالة سوء قناة إتصال المستخدم الأساسي. وفي حالة تحسن قناة إتصال المستخدم الأساسي يمكن للمستخدم الثانوي مشاركة المستخدم الأول في نفس التردد لنقل معلومات المستخدم الثانوي في نفس الوقت. تم عمل محاكاة للنموذج المقترح باستخدام برنامج الماتلاب ومقارنة النتائج للنموذج المقترح مع النماذج المشابهة الأخرى لإثبات التحسن فى أداء الشبكة.

تقسم الرسالة إلى خمس أبواب كالتالي :

في الباب الأول تم تقديم خلفية لموضوع الرسالة والمقصود بشبكات الراديو الإدراكي وكيفية إستخدامها ومكوناتها وأنواعها وكذلك تم توضيح الهدف من الرسالة وكيفية تنظيمها.

وقد تم في الباب الثاني عرض الدراسات السابقة فى مجال شبكات الراديو الإدراكي وأهم نقاط البحث والتي اشتملت علي أربع نقاط رئيسية في هذا الموضوع. وكذلك تم عرض الدراسات السابقة فى موضوع التعاون في إستخدام شبكات الراديو الإدراكي وإستبدال المصدر الثابت للقدرة الكهربية بخلايا الشحن مع الإستغلال الأمثل للطاقة فى حالة عدم وجود المصدر الثابت للتيار.

وقد تم في الباب الثالث عرض النموذج المقترح الأول الخاص بإمداد مستخدمي شبكات الراديو الإدراكي (المستخدم الرئيسي والثانوي) بخلايا شحن الطاقة يمكنها إعادة الشحن أثناء التشغيل والذي يهدف إلى الإستغلال الأمثل للطاقة . كذلك تم عمل التحليل الرياضي لأداء النموذج المقترح مع شرح درجة تعقيده والتقريبات الممكنة. وتم عمل مقارنة النتائج للنموذج المقترح مع النماذج المشابهة الأخرى لإثبات صحته.

وقد تم في الباب الرابع عرض الدراسات السابقة المختصة بتنظيم درجة التعاون بين المستخدم الثانوي والمستخدم الأساسي لشبكات الراديو الإدراكي. تم عرض المقترح الخاص بتنظيم درجة التعاون إعتمادا علي حالة قناة الإتصال للمستخدم الرئيسي مع السماح للمستخدم الثانوي باستخدام نفس التردد في نفس الوقت مع المستخدم الأساسي ولكن بقدرة منخفضة. تم عمل الفرض الرئيسي الخاص بطبيعة قناة الاتصال للمستخدم الرئيسي وباقي قنوات الإتصال. كذلك تم التحليل الرياضي للنموذج المقترح وعمل محاكاة للنموذج المقترح بإستخدام برنامج المحاكاة ماتلاب لمقارنة نتائج النموذج المقترح مع النماذج المشابهة الأخرى.

وقد تم في الباب الخامس عرض وتلخيص لنتائج الرسالة مع عرض الإقتراحات الخاصة بالأعمال المستقبلية التي يمكن إضافتها لموضوع الرسالة.

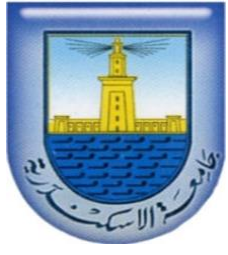

جامعة الاسكندرية

كلية الهندسة

قسم الهندسة الكهربية شعبة الإتصالات والإلكترونيات

# نماذج لتعاون شبكات الراديو الإدراكى

رسالة مقدمة

كإستكمال جزئي للحصول علي درجة الماجستير

من

**المهندس / رامي محمد عامر غانم عامر**

تحت إشراف

الأستاذ الدكتور / محمد عمرو مختار

الأستاذ الدكتور / عمرو أحمد الشريف

الدكتورة / هناء عبد العزيز ابراهيم

الإسكندرية

2016